\documentclass[useAMS,usenatbib]{mn2e}

\usepackage{epsfig}
\usepackage{graphicx}
\title[]{Constraining the mass distribution of galaxies using
  galaxy-galaxy lensing in clusters and in the field}

\author[Limousin et al.]{Marceau Limousin,$^1$ Jean-Paul Kneib$^{1,2}$ \& Priyamvada Natarajan$^3$ \\
$^1$ Observatoire Midi-Pyr\'en\'ees, UMR5572, 14 Avenue Edouard Belin, 31400 Toulouse, France.\\
$^2$ Caltech, Astronomy, MC105-24, Pasadena, CA 91125, USA.\\
$^3$ Dept. of Astronomy, Yale University, P.O. Box 208101, New Haven, CT 06250, USA.}
\date{}

\pagerange{\pageref{firstpage}--\pageref{lastpage}} \pubyear{2004}

\def\LaTeX{L\kern-.36em\raise.3ex\hbox{a}\kern-.15em
    T\kern-.1667em\lower.7ex\hbox{E}\kern-.125emX}

\begin{document}
\label{firstpage}
\maketitle

\begin{abstract}
We present a maximum-likelihood analysis of galaxy-galaxy lensing
effects in galaxy clusters and in the field. The aim is to determine
the accuracy and robustness of constraints that can be obtained on
galaxy halo properties in both environments - the high density cluster
and the low density field. This paper is theoretically motivated,
therefore, we work exclusively with simulated data 
(nevertheless defined to match observations) to study the accuracy with
which input parameters for mass distributions for galaxies can be
extracted. We model galaxies in the cluster and the field using a wide
range of mass profiles: the truncated pseudo isothermal elliptical
mass distribution, the Navarro, Frenk and White profile, and a Power
Law model with a core radius. We find that independent of the choice
of profile the mean mass of galaxies (of the order of
$10^{12}\,M_{\odot}$) can be estimated to within 15\% from
ground-based data and with an error of less than 10\% with space
observations. Additionally robust constraints can be obtained on the
mean slope of the mass profile. The two standard parameters that
characterise galaxy halo models, the central velocity dispersion and
the truncation radius can also be retrieved reliably from the
maximum-likelihood analysis. We find that there is an optimal scale
$R_{\rm max}$ which marks the boundary between lenses that effectively
contribute to the measured shear. Lenses beyond $R_{\rm max}$ in fact
dilute the shear signal. Furthermore, going beyond the usual 
formulation, we propose a re-parameterisation of the mass models that
allows us to put yet stronger constraints on the aperture mass of a
galaxy halo (with less than 10\% error). The gain in signal to noise
using space observations, expected for instance with the proposed SNAP
satellite compared to ground based data in terms of accuracy of
retrieving input parameters is highly significant. 
\end{abstract}

\begin{keywords}
Galaxies, Dark Matter Halos, Cosmology, Lensing.
\end{keywords}

\section{Introduction} 

Gravitational lensing has now become a popular tool to measure the
mass distribution of structure in the Universe on a range of
scales. Recently, there has been considerable progress in mapping the
mass distribution on relatively large scales using cosmic shear
\citep{refregier03}, and on cluster scales combining strong and weak
lensing features (\citealt{gavazzi03}; \citealt{kneib03}). On the
scale of individual galaxies as well, there has been much work done on
modeling and understanding multiple quasar systems
(\citealt{fassnacht}; \citealt{phillips}). In fact, in many cases it
has become clear that it is almost never a unique lens that is
responsible for the detected lensing and the presence of a nearby
galaxy, group or cluster along the line of sight plays an important
role in inducing the shear and amplification (\citealt{keeton};
\citealt{kneib00}; \citealt{moller02}). In other words, there are likely
no clean lines of sight and comprehensive modeling is needed to map the lensing
configuration accurately. Therefore the mass mapping problem is best
tackled using an ``inverse'' approach where the adopted method is to
model the distribution of matter around many lines of sight, and
optimise the mass distribution to match the observations as closely as
possible.

Analysing galaxy-galaxy lensing using maximum-likelihood lensing
techniques is an example of such a method. Indeed, the goal of
galaxy-galaxy lensing is to obtain constraints on the physical
parameters that characterise the dark matter halos of galaxies. This
is accomplished directly using lensing since the deformation in the
shapes of background galaxies produced by the foreground lenses
although weak is observationally detected statistically. The
difficulty is that multiple deflections frequently occur along the
line of sight, and therefore, nearby groups or clusters can have an
important effect yet again on the resultant distorsions. This
introduces a systematic bias in the mass obtained for the deflectors
when using simple models.

Galaxy-galaxy lensing work began with the first detection of the
signal from ground based data (Brainerd, Blandford \& Smail 1996,
hereafter BBS) and later with the {\it Hubble Space Telescope} (HST)
data \citep{Griffith96}. Maximum-likelihood techniques have been
developed by \cite{Rix}, \cite{NK96} and \cite{Geiger} to obtain
constraints on galaxy halo properties in clusters and in the
field. The results of these analyses suggest that galaxy halos in
clusters are significantly less massive but more compact compared to
galaxy halos around field galaxies of equivalent luminosity
(\citealt{Priya1}, \citealt{Priya2}). Besides, in the case of galaxy
halos in the field no clear edge is detected to the mass distribution
even on scales of the order of a few hundred kpc (\citealt{McKay};
\citealt{fisher}). Only two published studies to date by
Hoekstra (2003); Hoekstra, Yee \& Gladders (2004) have been able to
put an upper bound on the characteristic extension of a field halo at
about $290^{+139}_{-82}\,h^{-1}$ kpc and $185^{+30}_{-28}\,h^{-1}$
kpc, which are only marginally consistent with each other. Besides
these large values do not impose a stringent constraint for typical
galaxy mass distributions since at these typical radii the galaxy
density is only a few times above the mean density of the Universe.

Galaxy-galaxy lensing studies provide information on average
properties of the halo population, therefore the results depend on the
specific parameterised model chosen to fit the observational
data. From a purely observational point of view, the reliability of
the galaxy-galaxy lensing signal depends on the number density of
galaxies whose distorted shapes can be reliably measured, as well as
any additional constraints that can be added to the analysis, for
instance, redshifts of the lens galaxies, redshifts of the source
galaxies, galaxy type,
dynamical constraints, and the presence of larger scale structure like
groups or clusters in the vicinity.

Other methods to determine the masses of galaxies are generally based
on the dynamical properties of the luminous matter: measurement of the
rotation curve or velocity dispersion, study of the velocity field of
nearby objects like planetary nebulae, globular clusters, and
satellite galaxies. These dynamical methods are complementary to
lensing, but often probe much smaller scales. The study of the
velocity field around galaxies for instance, is generally limited to
local galaxies, however with the large spectroscopic surveys such as
2dF and SDSS, it is now possible to extend such analyses to larger
scales (\citealt{prada03}; \citealt{brainerd03}). Probing the dynamics
of stars in galaxies is limited to the inner regions when studying
high redshift galaxies. Therefore, at the present time there is
limited direct overlap between lensing and dynamical studies in terms
of scales probed. Although this situation is likely to change in the
very near future when large spectroscopic surveys of distant galaxies
such as the DEEP2 survey \citep{deep2}, the VVDS survey \citep{vvds}
or the z-COSMOS survey (http://webast.ast.obs-mip.fr/zCosmos) are
completed.

The inner slopes of density profiles provide a strong test of
structure formation in cold dark matter models and lensing provides an
unbiased way to estimate the slopes. \citet{treu04} and \citet{treu03}
have studied the slope of the mass distribution at small radii (on
scales ranging from a few to about 20 kpc) by combining dynamical
estimates and strong lensing constraints. They find that the mass
distribution profile is flatter than the singular isothermal sphere
(SIS) profile but steeper than the NFW profile. Therefore, there is
mounting evidence for the lack of cores (a constant density region) in
galaxies as well as in clusters.

The galaxy-galaxy lensing results from the Sloan Digital Sky Survey
have also provided (\citealt{McKay}; \citealt{sheldon03}) interesting
constraints on the distribution of light and dark matter in
galaxies. Mass and light trace seem to trace each other reasonably
well. The power of galaxy-galaxy lensing is that it provides a probe
of the gravitational potential of the halos of galaxies out to large
radii, where no other methods are viable for inter-mediate as well as
high redshift galaxies independent of the dynamical state of the
system. A similar approach combining dynamical estimates of the
central part of galaxies and galaxy-galaxy lensing is planned in the
future as part of the GEMS and COSMOS projects.

This paper is organised as follows: in section 2, we describe the
method adopted to model galaxy lenses, source galaxies, the
simulations performed to recover the \textsc{input} parameters of the
lenses and the calculation of the aperture mass. In section 3, we
present the results for three different classes of lens models
considered in this work. In section 4, we explore the results of
re-parameterising the models. Whenever necessary our results are
scaled to the currently preferred flat, low matter density
$\Lambda$CDM cosmology with $\Omega_M = 0.3, \ \Omega_\Lambda = 0.7$
and a Hubble constant $H_0 = 65 \ h_{65}$ kms$^{-1}$ Mpc$^{-1}$. In
such a cosmology, at $z=0.2$, $1''$ corresponds to $3.55 h_{65}^{-1}$ kpc.

\section{Galaxy-Galaxy Lensing}

We briefly review the basic principles of gravitational lensing of
distant galaxies before describing the mass distributions adopted to
model them. This section concludes with the presentation of our method
to recover the lensing galaxy parameters.

\subsection{Lensing equation}

The light rays emitted by a distant galaxy are distorted en-route to
us by the presence of mass concentrations along the line of sight.
The distorsion can produce strong effects like multiple images or arcs
if there is close alignment between the distant source and a
foreground source, but most of the time only a weak distorsion occurs
in the galaxy shape.

Let us consider the multiple lensing equation \citep{schneider92}.
For two lenses A and B, the lens equation becomes:
\begin{equation}
\vec{\beta}=\vec{\theta}-\vec{\alpha}_\mathrm{A} 
\frac{D_\mathrm{AS}}{D_\mathrm{OS}}
-\vec{\alpha}_\mathrm{B}\frac{D_\mathrm{BS}}{D_\mathrm{OS}}
\end{equation}
where $\vec{\beta}$ is the source position, $\vec{\alpha}_\mathrm{A}$
is the deflection due to the lens A, $\vec{\alpha}_\mathrm{B}$ is the
deflection due to the lens B and $D_\mathrm{AS}$, $D_\mathrm{BS}$ and
$D_\mathrm{OS}$ are the angular diameter distances between source
plane S and lens A, lens B and the observer respectively (note that we
must have $z_A<z_B<z_S$). The deflection angle
$\vec{\alpha}_\mathrm{X}$ due to the lens $X$ is proportional to the
angular distance $D_\mathrm{OX}$ between observer and lens $X$ and to
the gradient of the projected gravitational potential $\phi$ generated
by the lens $X$.

For a given background galaxy (i) and its associated lens (j), we can 
construct the amplification matrix $a_{ij}$, which provides the
mapping between the source plane and the image plane:
\begin{equation}
a_{ij}=
\left(
\begin{array}{cc}
1-\kappa^{ij}-\gamma_1^{ij} & -\gamma_2^{ij} \\
-\gamma_2^{ij} & 1-\kappa^{ij}+\gamma_1^{ij}
\end{array}
\right)
\end{equation}
where $\kappa$ is the convergence, and $\gamma_1$, $\gamma_2$ are the
two components of the shear. In the case of multiple deflections (more
than one lens contributing to the observed distortion), we will assume
that the total amplification matrix $a_i$ of the distant galaxy (i) is
equal to the sum of the individual contributions $a_{ij}$ from to each
lens:
\begin{equation}
a_i=\sum_{j} a_{ij}
\end{equation}
This assumption relies on the fact that we are in the weak lensing
regime and that the distance between the lenses is large compared to
the Einstein radius of each individual lens. For instance, in the
simulations performed in this paper, the separation between two lenses
is larger than 3", when a typical value for the Einstein radius is
about 1".

\subsection{Modeling the mass distribution of galaxies}

Lensing probes the two dimensional projected mass along the line of
sight, therefore, we deal with the two dimensional potential,
$\phi(R)$, resulting from the three dimensional density distribution
$\rho(r)$ projected onto the lens plane. The related projected surface
mass density $\Sigma$ is then given by:
\begin{equation}
\label{equ1}
4\pi\mathrm{G}\Sigma(R)=\nabla^2\phi(R)
\end{equation}

Moreover, we are interested in the two-dimensional projected mass inside 
radius $R$ (the aperture radius $R_{aper}$) defined as follows:
\begin{equation}
\label{equ2}
M_{aper}(R)=2\pi\int_0^R\Sigma(r)rdr
\end{equation}

In this paper, we study three different mass models (i) 
the two component pseudo-isothermal mass distribution (\emph{PIEMD})
\citep{stronglensing}, which is a more physically motivated mass profile than
the isothermal sphere profile (\emph{SIS}) but sharing the same profile slope
at intermediate radius; (ii) the \emph{NFW} \citep{nfw} profile and (iii) a 
Power Law profile with core radius (\emph{PL}). These enable us to
explore a wide range of mass distributions and reveal the important 
parameters that lensing is sensitive to.

\subsubsection{PIEMD profile}

The density distribution for this model is given by:
\begin{equation}
\label{rhoPIEMD}
\rho(r)=\frac{\rho_0}{(1+r^2/r_{core}^2)(1+r^2/r_{cut}^2)}
\end{equation}
with the core radius $r_{core}$ of the order of 100 pc, and a
truncation radius $r_{cut}$. We also introduce a shape parameter
$a=r_{cut}/r_{core}$.  In the centre,
$\rho\simeq\rho_0/(1+r^2/r_{core}^2)$ which describes a core with
central density $\rho_0$. The transition region ($r_{core}<r<r_{cut}$)
is isothermal, with $\rho\simeq r^{-2}$. In the outer parts, the
density falls off as $\rho\simeq r^{-4}$, as is usually required for
models of elliptical galaxies. Fig.~\ref{fig1} illustrates this
behaviour. These models have been successfully used by Natarajan et al.
(1998, 2002) to fit observed early-type galaxies in cluster lenses.

Integrating equation \ref{rhoPIEMD}, we obtain the 2 dimensional
surface mass density distribution:\\
\begin{equation}
\label{2DPIEMD}
\Sigma(R)=\frac{\sigma_0^2 r_{cut} }{2\mathrm{G}(r_{cut}-r_{core})}(\frac{1}{\sqrt{r_{core}^2+R^2}}-\frac{1}{\sqrt{r_{cut}^2+R^2}})
\end{equation}
where $ \sigma_0 $ is the central velocity dispersion for a circular
potential related to $\rho_0$ by the following relation:
\begin{equation}
\label{rhodesigma}
\rho_0=\frac{\sigma_0^2}{2\pi\mathrm{G}}(\frac{r_{cut}+r_{core}}{r_{core}^2 r_{cut}})
\end{equation}
It can be shown that for a vanishing core radius, the
surface mass profile obtained above becomes identical to the surface
mass profile used by BBS for modeling
galaxy-galaxy lensing. The enclosed two dimensional aperture mass
interior to radius $R$ is:
\begin{equation}
\label{MPIEMD}
M_{aper}(R)=\frac{\pi r_{cut} \sigma_0^2}{\mathrm{G}}
(1 - {\sqrt{r_{cut}^2+R^2}-\sqrt{r_{core}^2+R^2} \over r_{cut}-r_{core}})
\end{equation}
and the total mass of such a model is finite and is given by:
\begin{equation}
M_{tot}=\frac{\pi \sigma_0^2}{\mathrm{G}}\frac{r_{cut}^2}{r_{cut}+r_{core}}
\simeq \frac{\pi \sigma_0^2 r_{cut}}{\mathrm{G}}
\end{equation}

Fig.~\ref{fig1} shows the behaviour of $M_{aper}$ as a
function of the aperture radius $R_{aper}$ and of $\rho(r)$ for such a
profile, with $\sigma_0=\,220\,\mathrm{kms}^{-1}$, $r_{core}=0.1"$, $r_{cut}=\,30"$.
$\rho(r)$ is also shown, normalised to the critical density of the Universe
 $\rho_{\mathrm{crit}}$, where $\rho_{\mathrm{crit}}=3H_0^2/8 \pi G$.

\subsubsection{NFW profile}

The NFW density profile 
\citep{nfw} provides the best-fit to the halos that
form in N-body simulations of collisionless dark matter. 
In fact the NFW profile reproduces with good
accuracy the radial distribution of structures in these
simulations over 9 orders of magnitude in mass (from the scale 
of globular clusters to that of massive galaxy clusters). Since it is 
thought that matter in the Universe is dominated by a form of
dissipationless cold dark matter, this ``universal profile'' offers an
interesting and natural way of describing mass concentrations. 
The density distribution of the NFW profile is given by:
\begin{equation}
\label{rhoNFW}
\rho(r) =  \frac{\rho_s}{(r/r_{s})(1+r/r_{s})^{2}}
\end{equation}
where $\rho_s$ is a characteristic density. It is possible to
parameterise this model in terms of $M_{200}$, which is the mass
contained in a radius $r_{200}$ where the criterion
$\overline{\rho}$=200$\rho_{\mathrm{crit}}$ holds, and $\delta_c$ the density
contrast (or equivalently $c=r_{200}/r_s$, the concentration parameter). We have
the following relations between the two parameterisations:
\begin{equation}
\rho_s=\delta_c\rho_c,\,\,M_{200}=\frac{800}{3}\pi r_{200}^3
\rho_c,\,\,\\
\delta_c=\frac{200}{3}\frac{c^3}{ln(1+c)-\frac{c}{1+c}}
\end{equation}
The properties of the projected quantities depends on the ratio
$r/r_{s}$, so it is useful to introduce the
dimensionless radial coordinate, $x=r/r_{s}=R/r_s$. Moreover, the
velocity dispersion $\sigma(r)$ of this potential, computed with the
Jeans equation assuming an isotropic velocity distribution, gives an
unrealistic central velocity dispersion ($\sigma(0)=0$). In order to
compare the NFW potential with other potentials used to model lenses, 
we define a characteristic velocity $\sigma_s$ as follows:
\begin{equation}
\label{sigmaNFW}
\sigma^2_s=\frac{4}{3}\mathrm{G}r_{s}^2\rho_s
\end{equation}
The surface mass density for the NFW is given by:
\begin{equation}
\label{2DNFW}
\Sigma(x)=\int_{-\infty}^{+\infty} \rho(r_{s},x,z) dz=2 \rho_s r_{s} F(x)
\end{equation}
with
\begin{equation}
F(x)= \left\{\begin{array}{ll}
\frac{1}{x^2-1}(1-\frac{1}{\sqrt{1-x^2}}\textrm{argch}\frac{1}{x}) & \textrm{(x$<$1)} \\
\frac{1}{3}  & \textrm{(x=1)} \\
\frac{1}{x^2-1}(1-\frac{1}{\sqrt{x^2-1}}\arccos\frac{1}{x}) & \textrm{(x$>$1)}
\end{array}
\right.
\end{equation}
and the two dimensional aperture mass $M_{\rm aper}$ contained within the dimensionless
radius x is \citep{bartelmann}:

\begin{equation}
\label{MNFW}
M_{\rm aper}(R)=\frac{3\pi\sigma^2_s r_{s}}{2\mathrm{G}}g(x)
\end{equation}
with
\begin{equation}
g(x)= \left\{\begin{array}{ll}
\ln\frac{x}{2}+\frac{1}{\sqrt{1-x^2}}\textrm{argch}\frac{1}{x} & \textrm{(x$<$1)} \\
\ 1+\ln(\frac{1}{2})  & \textrm{(x=1)} \\
\ln\frac{x}{2}+\frac{1}{\sqrt{x^2-1}}\arccos\frac{1}{x} & \textrm{(x$>$1)} 
\end{array}
\right.
\end{equation}
and the mass $M_{200}$ can be written as a function of $\sigma_s$, $r_{200}$ 
and $c$:
\begin{equation}
M_{200}=200 \pi \frac{c^2}{\delta_c}\frac{\sigma_s^2 r_{200}}{\mathrm{G}}
\end{equation}

Fig.~\ref{fig1} shows the behaviour of $M_{aper}$ as a function of
the aperture radius $R_{aper}$ and of $\rho(r)$,
with $\sigma_s$ = 225 kms$^{-1}$ and  $r_{s}$ = 3".  This profile has a concentration
parameter $c=r_{200}/r_s\simeq\,12$, a typical value for a
galaxy, and a projected mass inside $r_{200}$: $M_{200}$ of
$\sim\,3 \times\,10^{12}\,M_{\sun}$ .

\subsubsection{Power Law profile with a core}

Another simple model to describe the mass distribution of a galaxy is
a power-law model (PL) with a core. In a CDM dominated hierarchical
structure formation scenario, mass profiles are expected to be
independent of the mass scale, therefore a power-law profile is of
interest.  The PL mass distribution has three parameters: a core
radius $r_{core}$ of the order a kpc for an average galaxy, a central
velocity dispersion, $\sigma_0$ measured in $(\mathrm{km s}^{-1})$ and an exponent ($\alpha$)
which defines the gradient of the mass distribution. The
three-dimensional density profile is (Kneib 1993):
\begin{equation}
\label{sigmaPL}
\rho(r)=\rho_0 \frac{1+\frac{1-2\alpha}{3} (\frac{r}{r_{core}})^2}{(1+(\frac{r}{r_{core}})^2)^{2+\alpha}}
\end{equation}
Introducing $x=\frac{r}{r_{core}}$, the density profile falls off as 
$\rho\simeq x^{-2(1+\alpha)}$. Note that the case $\alpha$=0 corresponds
to an isothermal sphere with a core radius, and  
$\alpha>$0 defines density profiles steeper than an isothermal sphere
with a core radius.
The relation between $\rho_0$ and $\sigma_0$ is given by:
\begin{equation}
\label{rhodesigmapl}
\rho_0=\frac{\sigma_0^2}{r_{core}^2}\frac{9(1-2\alpha)}{4\pi\mathrm{G}}
\end{equation}
The surface mass density is:
\begin{equation}
\label{2DPL}
\Sigma(x)=\Sigma_0\frac{1+\frac{1-2\alpha}{2}x^2}{(1+x^2)^{3/2+\alpha}}
\end{equation}
and the two dimensional aperture mass contained within the
dimensionless radius x is:
\begin{equation}
\label{M_PL}
M_{aper}(R)=\frac{3(1+2\alpha)\sigma_0^2 r_{core}}{\mathrm{G}} \frac{x^2+x^4}{(1+x^2)^{3/2+\alpha}} I_{1+\alpha}
\end{equation}
with
\begin{equation}
I_{1+\alpha}=\int_0^{\infty}\frac{dx}{(1+x^2)^{1+\alpha}}
\end{equation}

\begin{figure}
\includegraphics[height=7cm,width=7cm]{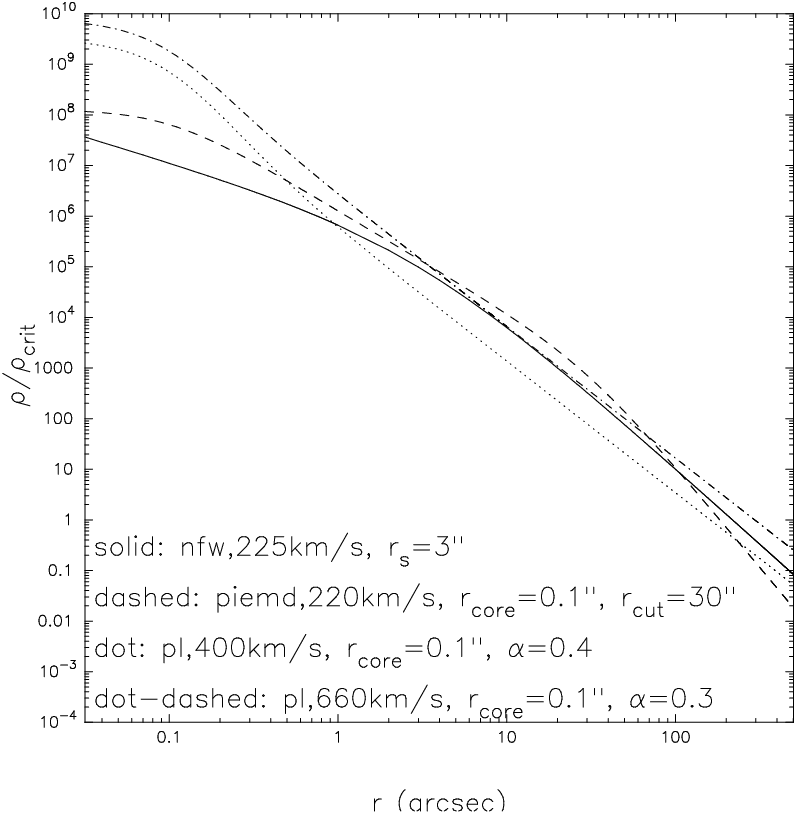}
\includegraphics[height=7cm,width=7cm]{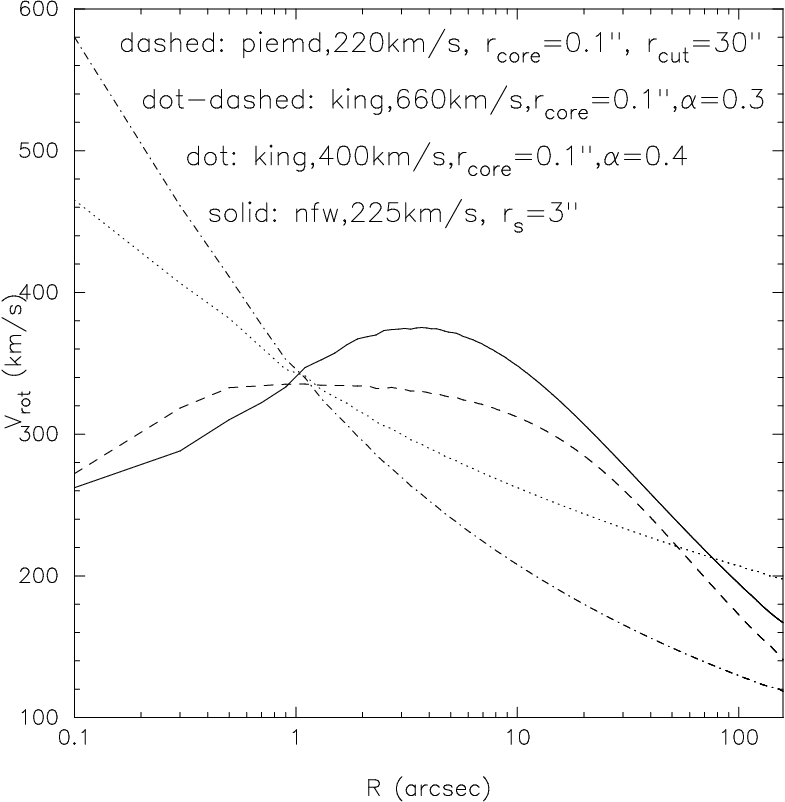}
\includegraphics[height=7cm,width=7cm]{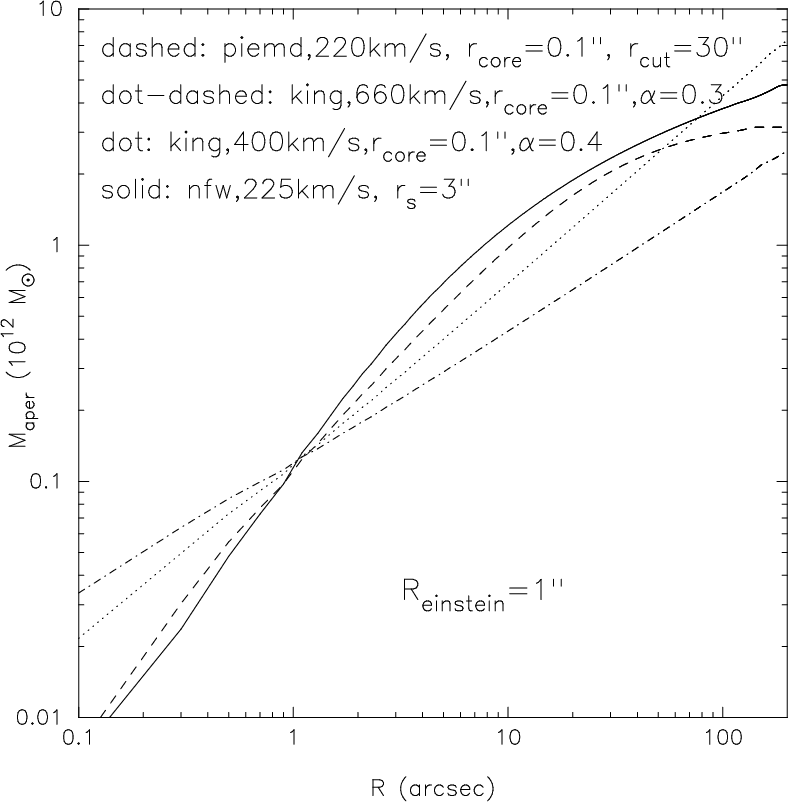}
\caption{The density profile $\rho(r)$, the rotation velocity $V_{rot}(R)$ and the aperture mass $M_{aper}(R)$ for the 3 different mass profiles studied in this work. For each of these model profiles the relevant parameter choices are shown in the figure.}
\label{fig1}
\end{figure}

Fig.~\ref{fig1} shows the behaviour of $\rho(r)$ and
$M_{aper}(r)$, for such a profile, with $\sigma=400\,\mathrm{kms}^{-1}$ and $\alpha=0.4$ (in this case, $I_{1.4}$=1.06).

\subsection{Comparing these profiles}

Beginning with the Boltzmann equation to describe the behaviour of the
cold dark matter collisionless particles that constitute a galaxy
halo, the Jeans Equation for a spherical potential and for an
isotropic velocity distribution ($\sigma\equiv \sigma_r$) is given by:
\begin{equation}
\frac{1}{\rho}\,\frac{\mathrm{d}(\sigma^2\rho)}{\mathrm{d}r}=
-\frac{\mathrm{d}\Phi}{\mathrm{d}r}
\label{Jeans}
\end{equation}
where $\Phi$ is the three dimensional potential. Considering the limit
at +$\infty$, wherein $\sigma(r)\to0$ and $\rho(r)\to\quad$0 to
perform the integration, we have:
\begin{equation}
\sigma^2(r)=-\frac{1}{\rho(r)} \int_r^{+\infty} \rho(r')\frac{d\Phi(r')}{dr'}dr'
\label{sigma}
\end{equation} 
Solving this equation, does not lead generally
to a simple analytical expression for the velocity dispersion.

Hence, one usually uses the rotation velocity defined as:
\begin{equation}
V_{rot}^2(R)=\frac{\mathrm{G} M_{aper}(R)}{R}
\end{equation}

Fig.~\ref{fig1} shows the behaviour of $V_{rot} (R)$ for the 
three different mass profiles studied here.

It can be shown that for any spherically symmetric profile, the mass
inside the Einstein radius $R_E$ is proportional to $R_E^2$, so that
profiles for which $R_E$ is constant can be compared.  It is
easy to show ({\it e.g.} Kneib 1993) that:
\begin{equation}
M_{aper}(R_E)=\pi\Sigma_{\mathrm{crit}}\,R_E^2
\end{equation}

\begin{figure}
\begin{center}
\includegraphics[height=6cm,width=6cm]{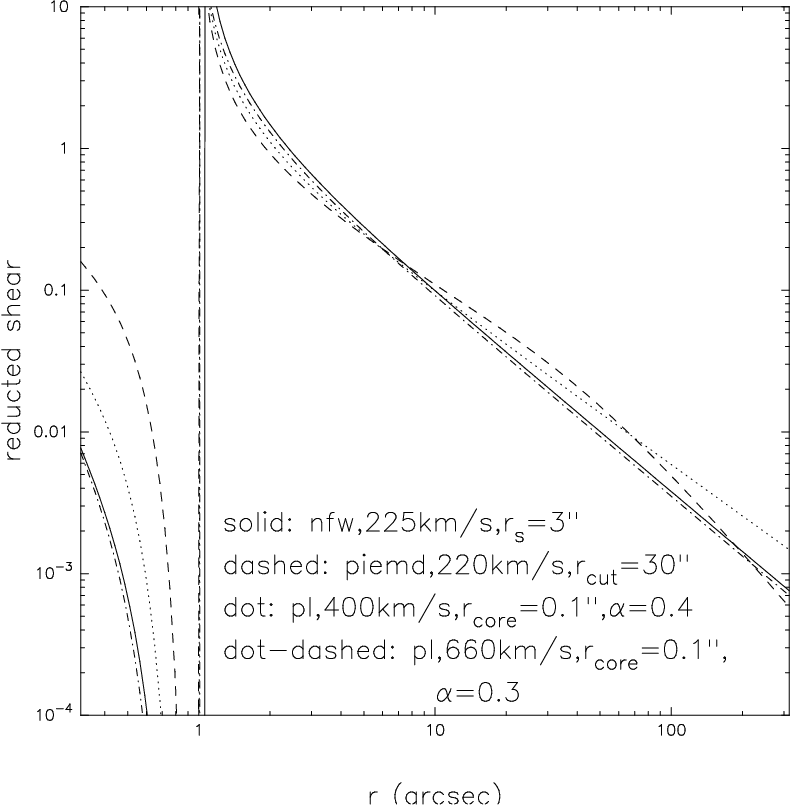}
\includegraphics[height=3cm,width=6cm]{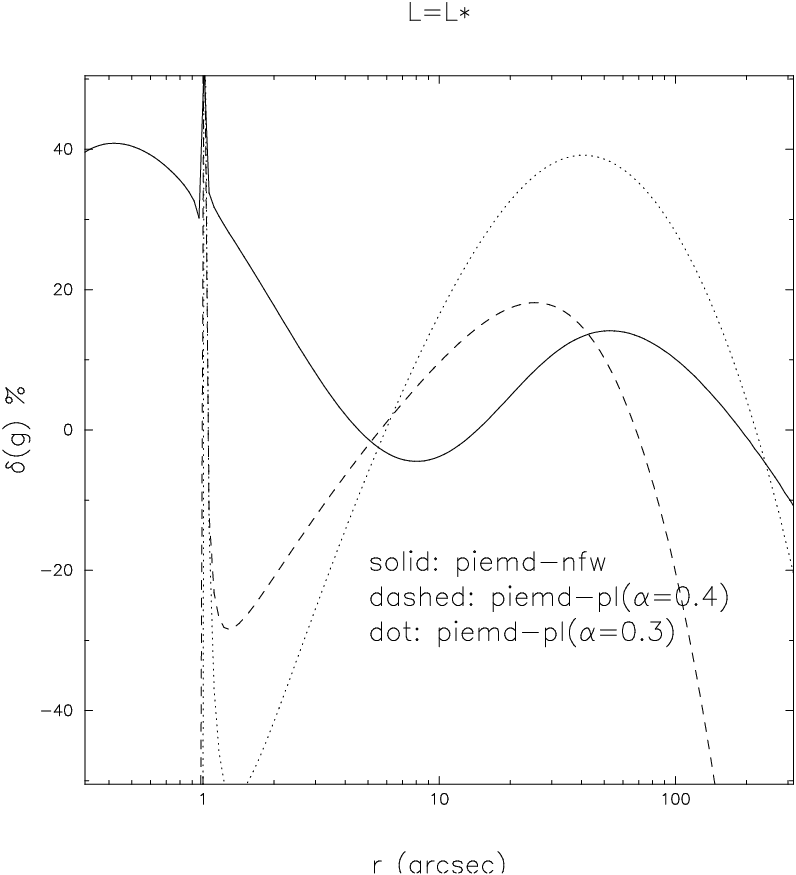}
\end{center}
\caption{\label{fig2} The reduced shears ($\frac{\gamma}{1-\kappa}$) for the 3
different models for which the Einstein radius $R_E=1"$ (upper panel), and
differences between them expressed as a percentage where we have used
the PIEMD as the reference profile (lower panel).}
\end{figure}

Thus we adjust the parameters of the different mass profiles in order to 
have the same Einstein radius, and therefore the same mass within the Einstein
radius.  The results are 
illustrated in Fig.~\ref{fig2}; this plot allows us to rescale the 
velocity dispersions derived for each profile.
We can immediately see from this plot that for $r>4"$, the shears
computed from the three profiles are comparable. 
Note that in the case of the PL profile there is
a strong dependence of the velocity dispersion on the value of the 
exponent $\alpha$. For example, the PL profile with
$\sigma_0=660\,\mathrm{kms}^{-1}$ and an exponent of 0.3 induces the same shear
as one with $\sigma_0=400\,\mathrm{kms}^{-1}$ and an exponent of
0.4. Therefore, there is degeneracy between the value of $\sigma_0$
and $\alpha$ for the PL profile.  In order to
illustrate the behaviour of the PL profile, we include this latter
profile in Fig.~\ref{fig1} and \ref{fig2}.

\section{Simulating galaxy-galaxy lensing}

\subsection{Measurement of background galaxy shapes}

We study in detail the lensing effects in two observational scenarios:
(i) a ground based survey using a wide field camera and (ii) data from
space-based observations. Ground based data is characterised by the
following image quality: seeing of about 0.6-0.8 arcsec, and a galaxy
number density of 20-40 galaxies per square arcminute, of which only
50-70\% can generally be used to measure their shapes reliably. The
above estimates were obtained from two hours of observation in the R
band with the CFH12k camera with a field of view of 44x28 square
arcminutes.

Space observations have a significantly better image quality, with a
PSF of about 0.1 to 0.15 arcsec, and a galaxy number density of about
40-100 galaxies per square arcminute (SNAP mission sensitivities - see
\citet{snap1}), of which about 80\%. can be used in a weak lensing
study since their shapes can be measured to the requisite degree of
precision. Half an hour of observation in the R band of A2218 with the
HST have provided the above estimates.  

The measurement of shapes of lensed background galaxies is made
using the second moment of the intensity of their light distribution.
The quantity that is extracted for each galaxy is its complex ellipticity, $e$ defined as
$e=e_1 + ie_2$. The magnitude of the ellipticity is $e=\sqrt{e_1^2 + e_2^2}$,
the mean ellipticity $m=\frac{\sum_i e_i}{N}$, the dispersion of
the ellipticity is equal to the square root of the variance, defined
by $var=\frac{\sum_i (e_i-m)^2}{N}$, where N is the number of
objects.

\subsection{Scaling the mass distribution from the light distribution}

The foreground lenses are described by a mass profile with known input
parameters. The parameters used to describe the different lenses are
scaled as a function of luminosity. The scaling relation for
$\sigma_0$ assumes that mass traces light, and its origin resides in
the Tully-Fisher or Faber-Jackson relations.  The scaling relation for
the radial parameter assumes that the mass-to-light ratio is constant
for all galaxies.  Note that they are other possible scaling
relations, and that in principle we can test them with lensing.

\subsubsection{PIEMD profile}

We have for this profile:
\begin{equation}
\sigma_0=\sigma_0^* (\frac{L}{L^*})^{\frac{1}{4}}\quad  \& \quad  
r_{cut}=r^*_{cut} (\frac{L}{L^*})^{\frac{1}{2}}
\end{equation}
The parameter $r_{core}$ is kept fixed at 0.1" -  a fairly typical
value for a galaxy.

From equation (13), we can scale the total mass with the luminosity as:
\begin{equation}
M_{tot}=\frac{\pi \sigma_0^2 r_{cut}}{\mathrm{G}}=
\frac{\pi \sigma_0^{* 2} r_{cut}^*}{\mathrm{G}} (\frac{L}{L^*})^{3/4}
\end{equation}

\subsubsection{NFW profile}

Similar to the PIEMD profile, we have:
\begin{equation}
\sigma_0=\sigma_0^* (\frac{L}{L^*})^{\frac{1}{4}}\quad  \& \quad  
r_{s}=r^*_{s} (\frac{L}{L^*})^{\frac{1}{2}}
\end{equation}

\subsubsection{PL profile}

For the PL profile:

\begin{equation}
\sigma_0=\sigma_0^* (\frac{L}{L^*})^{\frac{1}{4}} \quad  \& \quad
r_{core}=r^*_{core} (\frac{L}{L^*})^{\frac{1}{2}}
\end{equation}

\begin{figure}
\begin{center}
\includegraphics[height=6cm,width=6cm]{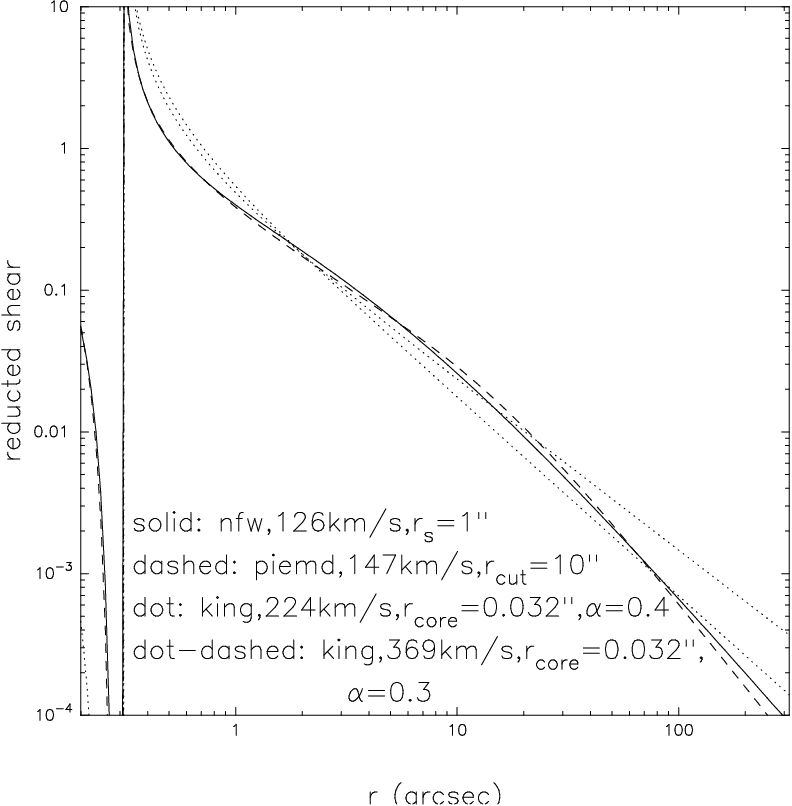}
\includegraphics[height=3cm,width=6cm]{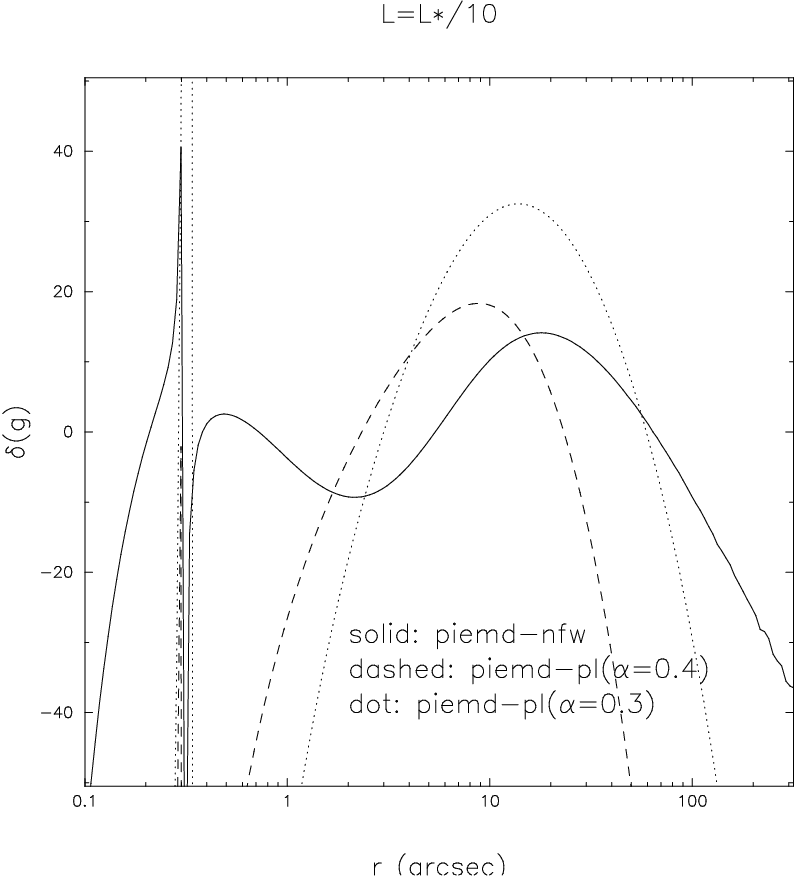}
\includegraphics[height=6cm,width=6cm]{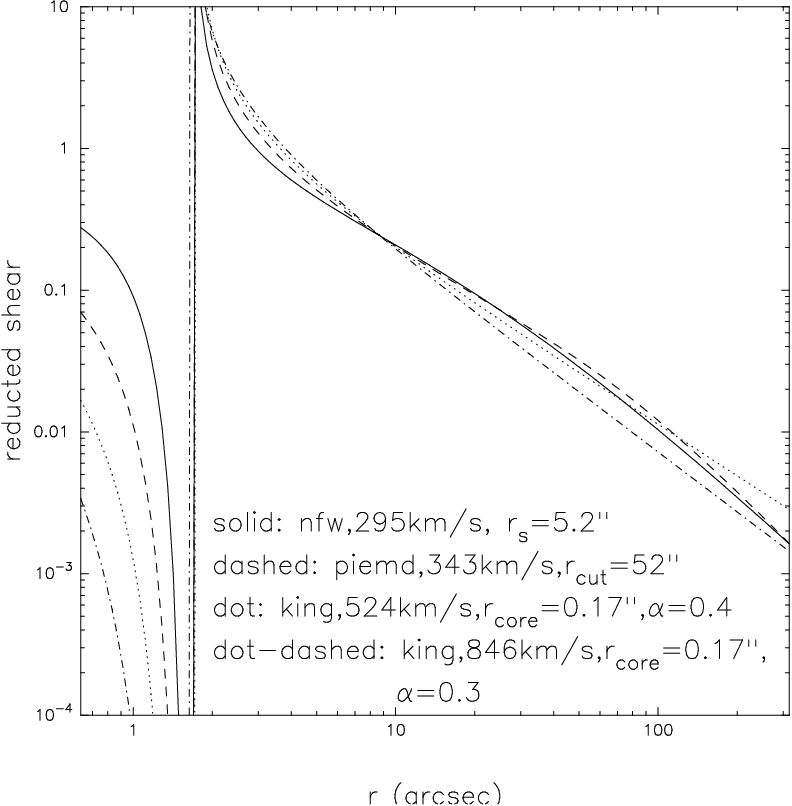}
\includegraphics[height=3cm,width=6cm]{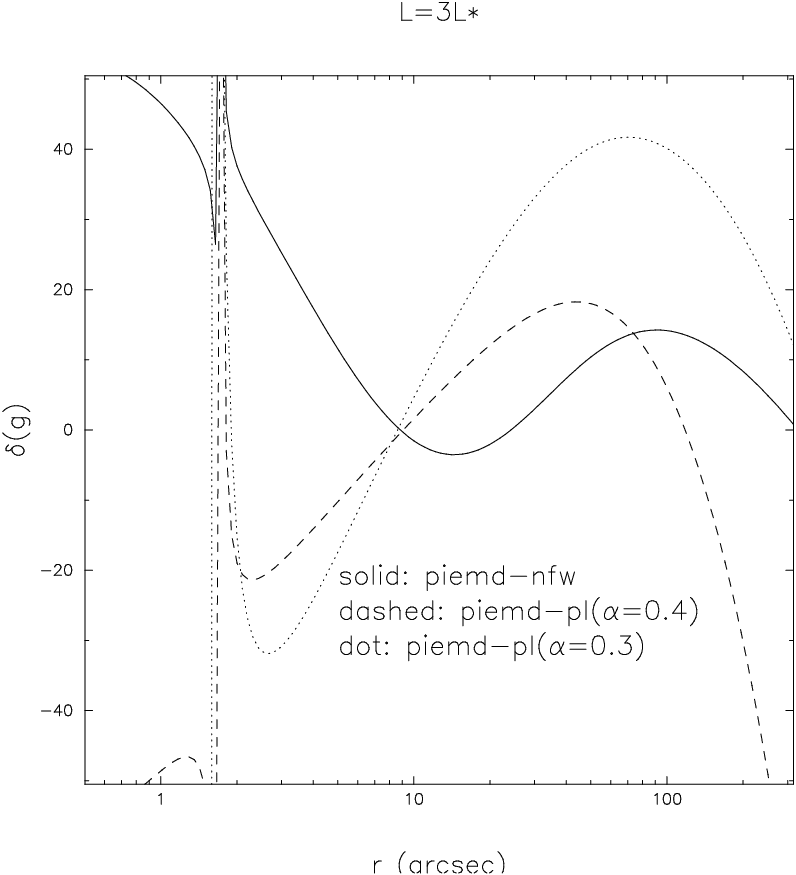}
\end{center}
\caption{
The reduced shears ($\frac{\gamma}{1-\kappa}$) for each
profile (Panels 1 and 3) and differences between them expressed in
percentages (Panel 2 and 4), for $L=L^*$/10 and $L=3L^*$ respectively,
where once again the PIEMD is the reference profile.}
\label{fig3}
\end{figure}

In order to illustrate the coherence of these scaling laws, we show
the shear profiles obtained for a typical faint ($L=L*/10$) and bright
($L=3L*$) galaxy (see Fig.~\ref{fig3}).

\subsection{Background galaxies}

\begin{figure*}
\begin{center}
\includegraphics[height=5cm,width=5cm]{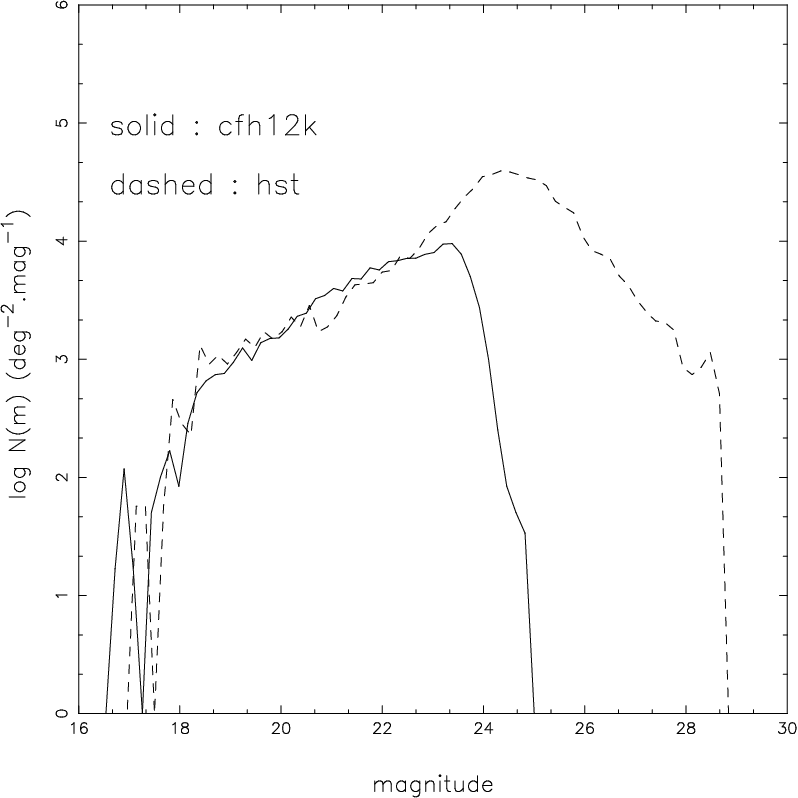}
\includegraphics[height=5cm,width=5cm]{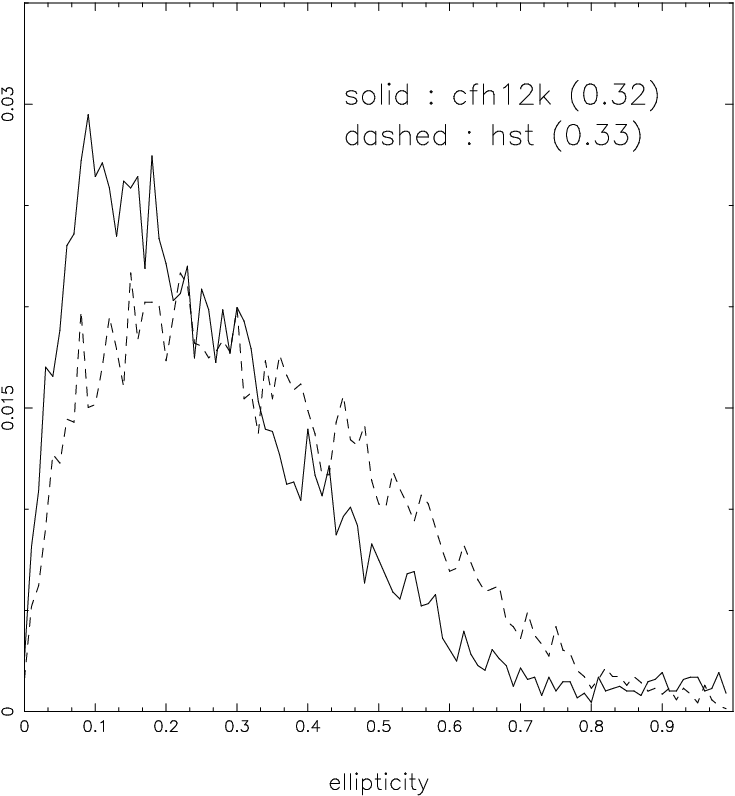}
\includegraphics[height=5cm,width=5cm]{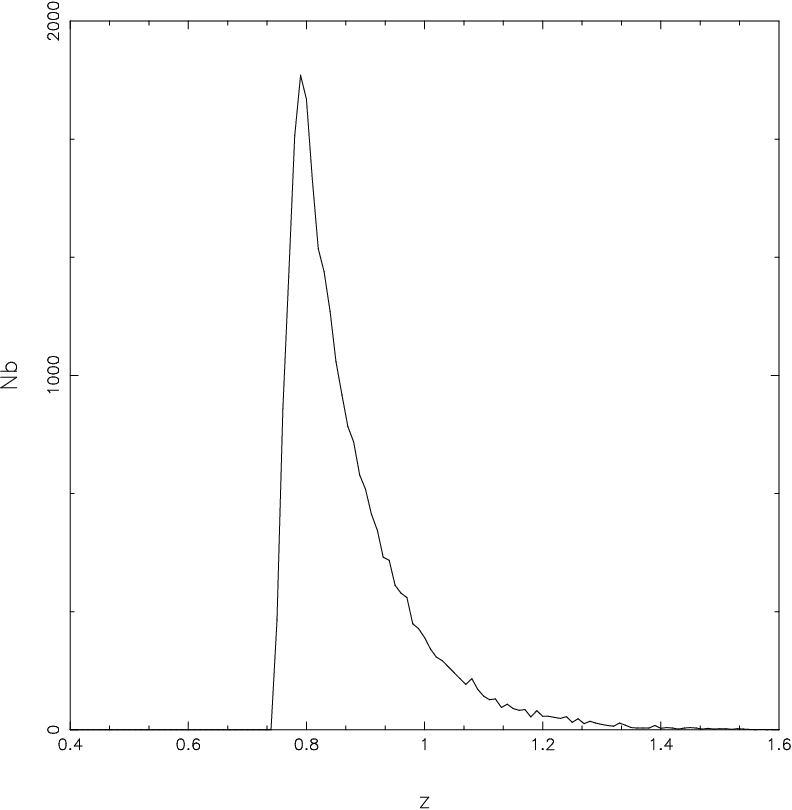}
\caption{Galaxy counts from CFH12k and HST data (Left panel), and the
  ellipticity distributions for the CFH12k and HST data (Center panel). The value in parentheses is the mean value of the ellipticity in each
  case. Right panel: the redshift distribution of the background population 
  in our simulations.}
\label{fig4}
\end{center}
\end{figure*}

The way we simulate the background source population is the same 
for the 2 cases when the lenses belong to a cluster versus when
they are field galaxies:
\begin{itemize}
\item they are allocated random positions 
\item number counts are generated in consonance with galaxy counts
  typical for a 2 hour integration time in the R-band. 
The magnitudes are assigned by drawing the number count observed with 
the \emph{CFHT}
\item the shapes are assigned by drawing the ellipticity from a
  gaussian distribution similar to the observed \emph{CFHT}
  ellipticity  distribution (see Fig.~\ref{fig4})
\item Redshift distribution: We use the R-band to define 
the number counts of 
galaxies and use the HDF prescription in terms of the mean redshift
per magnitude bin, and the same redshift distribution as BBS. 
\end{itemize}

\subsection{Lens galaxies}

The way we simulate the foreground lenses galaxies is different in the
case of galaxies inside the cluster and in the field.

\subsubsection{Cluster Galaxies}

We put the individual lenses constituting a cluster at a redshift of
0.2, and model it as a superposition of large-scale smooth cluster component and a few clumps.
In order to obtain a better match to the real data, the positions and the
magnitudes of the foreground cluster galaxies are drawn
from the positions and the magnitudes of ellipticals 
in the cluster A1689 at  $z = 0.18$.

\subsubsection{Field Galaxies}

For simplicity, the lenses are randomly distributed in position and uniformly
distributed in a redshift range from 0.2 to 0.5.
This distribution is a crude approximation of reality.

\section{Maximum Likelihood analysis}

\subsection{Methodology}

Using the foreground cluster and field galaxies as the lens for the
sheet of generated background galaxies, we use LENSTOOL to solve the
lensing equation and produce a catalogue of lensed background galaxies.
This catalogue contains the following information for each lensed
object: the position, the shape parameters, and the redshift. Then
this catalogue is processed through a numerical code that retrieves the 
input parameters of the lenses using a maximum-likelihood method as
proposed by \cite{Rix} and as implemented by
Natarajan \& Kneib (1997). For each image ($i$), given a mass model for 
the foreground lensing galaxies, we can compute the amplification matrix $a_i$
as a sum of the contribution from all the foreground galaxies $j ; z_j<z_i$ that lie
within a circle of inner radius $R_{min}$, and outer radius $R_{max}$
measured from the centre of the image ($i$):
\begin{equation}
a_i=\sum_{\begin{array}{c} {z_j<z_i} \\ d(i,j)<R_{max}\end{array}} a_{ij}
\end{equation}
The total shear experienced by a background galaxy $\gamma_i$ can be
obtained  by summing the contributions from all the foreground galaxies $j ; z_j<z_i$ that lie
within an annulus with inner and outer radii respectively at $R_{min}$
and $R_{max}$.

Given the observed ellipticity $\vec{\varepsilon^i_{obs}}$ 
(defined as $\varepsilon=(a-b)/(a+b)$)
and the associated amplification
matrix $a_i$, we are able to retrieve the intrinsic 
ellipticity $\vec{\varepsilon_i^s}$ of the source before lensing:
\begin{equation}
\vec{\varepsilon_i^s}=F(\vec{\varepsilon^i_{obs}}, a_i)
\end{equation}
In the weak lensing regime, this relation can be simplified as:
\begin{equation}
\vec{\varepsilon^s_i} =\vec{\varepsilon^i_{obs}} + \vec{\gamma_i}
\end{equation}
In order to assign a likelihood to the parameters used to describe the lensing
galaxies, we use 
$P^s$, the ellipticity probability distribution in the absence of
lensing.  Repeating this procedure for each image in the catalogue, we construct the
likelihood function:
\begin{equation}
\label{L}
\mathcal{L}=\prod_i P^s(\vec{\varepsilon^s_i})
\end{equation}
which is a function of the parameters used to define the mass models
of the lenses. For each pair of chosen parameters, we can compute a 
likelihood function.  The larger the likelihood function, the more
accurate the retrieved parameters used to describe the lenses.
The inversion from the observed ellipticity 
to the intrinsic ellipticity is fully analytic and takes into account 
all the non-linearities arising in the strong lensing regime, which may occur 
in the configuration with a cluster component.  

The likelihood function for the parametric mass model for the lenses does
have interesting convergence properties. The likelihood surface is
topologically complex since the degeneracies in the problem tend to
produce several maxima. However, the convergence in both the cluster
lens case and field lens case is driven essentially by the width
of the intrinsic ellipticity distribution of the sources. 
The degeneracies in this scheme are the generic ones that plague
all lensing analyses, the mass-sheet degeneracy (essentially the
addition of a constant sheet of mass to the lens plane does not
produce any discernable effect on the lensing of the background
sources), the shot noise due to the finite number of sampled
background sources and the details of the truncation of the mass
profile of the lenses. The mass sheet degeneracy cannot be
circumvented since we are necessarily measuring both the magnification
$\kappa$ and the shear $\gamma$ from the same data points. Note that
these are implicitly needed in computing the amplification matrix.
Shot noise is more of a limitation in ground-based surveys when the
number count of background galaxies is sparser compared to the space
based data despite the fact that lensing helps by magnifying fainter
sources that normally would not make it into a magnitude limited
survey. The details of the mass profile, and in fact, the prescription
used to truncate the mass at large radii influences the likelihood
results. Since in galaxy-galaxy lensing (both in the field and
interior to a cluster) we are most sensitive to the mass enclosed
within an aperture and are less sensitive to the density profile
in the inner regions the truncation of the mass distribution at
large radii drives the convergence of the likelihood function.
This can be clearly seen in the plots where the influence of
$R_{\rm max}$ shows up directly in the likelihood contours (see
Fig.~\ref{fig5}). Note that the parameters chosen to characterise the
mass model: the central velocity dispersion and the truncation
radius also contribute to the noise in the likelihood,
thereby pointing to more efficient re-parameterisations that
we also explore in the final section of this paper.

\subsection{Cluster weak lensing mass estimates}

We construct a composite mass model for the cluster by superposing a
large-scale smooth mass component and individual galaxies. As a first
guess for the smooth mass model we use the averaged shear field
obtained by simply binning up the shear in radial bins from the centre
outward. This is a prior in the analysis that gets modified with every
iteration once the clumps are added to the model. We simulate the
deformations induced by a clump with known parameters (which can be
easily derived from a weak lensing analysis for example). Then we add
in the individual cluster galaxies and derive the shear of this
composite system, which turns out to be larger than the shear for the
clump alone as expected: this implies that we need to simultaneously
modify the smooth component and the clumps during the optimisation
process. In massive lensing clusters, about 10\% of the total mass is
associated with the individual galaxies \citep{Priya2}. This
large scale clump is described by a PIEMD profile with the following
parameters: $\sigma_0=\,1070\,\mathrm{kms}^{-1}$, $r_{cut}=\,930\,kpc$ and
$r_{core}=\,60\,kpc$. This gives a total integrated mass of 7.3
10$^{14}$M$_{\sun}$. The mass we partition to  galaxies is of the order of
7.3 10$^{13}$M$_{\sun}$. How do we modify the large-scale clump's  parameters
in order to match the shear field? We find that the
velocity dispersion of the large-scale clump needs to reduced by about
5\%, keeping the others parameters fixed to accomodate the clumpiness.

\section{Results}

We present the results obtained for the simulated data set, for the
PIEMD, NFW and truncated power law (PL) models for two different
configurations (Fig.~\ref{fig9}, \ref{fig10} and \ref{fig11} at the
end of the paper). The points mark the value of the INPUT parameters
used in order to generate the simulated catalogue, and the cross
stands for the value of the OUTPUT parameters as estimated from the
maximum likelihood analysis.  We will refer to the different plots by
assigning them a number, the first one is the upper left plot, and the
last one (plot number 9) is the lower right plot.  The first plot
shows the reference field situation: 25000 elliptical sources in a
field of 26$\times$26 arcmin$^{2}$, which translates into a number
density of about 35 galaxies per arcmin$^{2}$. Then the following
plots (2, 3, 4) show the results obtained with 25000 circular sources,
then with 40000 elliptical sources mimicking typical space
observations with a density of 60 galaxies per arcmin$^{2}$, then with
12500 elliptical sources, corresponding to the ground based
configuration with a galaxy density of 17 per arcmin$^{2}$, in a field
configuration.  Note that in the case of circular sources, we do not
have to deal with the intrinsic ellipticity noise: the detection is
therefore improved and the contours are tighter.  The plots number 5
and 6 demonstrate the effect of the unknown redshift distribution for
the background sources, in a field configuration.  The last row
represents the cluster configuration: the 'standard' configuration
(plot 7), then configurations where an uncertainty on the cluster has
been introduced (plot 8 and 9).  for each plot, the contours represent
the 3$\sigma$,4$\sigma$,5$\sigma$ confidence levels, and along the
dotted lines, the mass within a projected radius $R_{aper}\,=\,100$ kpc is
constant, equal to the value indicated on the plot.

\subsection{The number of background lenses}

For each profile, in the field configuration, we explore the influence
of the background density on the detection. The 'standard'
configuration has 25000 background sources (a density of 35 galaxies
per arcmin$^{2}$), we explore what happens when we increase this
number to 40000 (60 per arcmin$^{2}$), or reduce it to 12500 (17 per
arcmin$^{2}$). 17 galaxies per square arcminute corresponds to simulating
ground based survey data, whereas 60 galaxies per square
arcminute corresponds to the space based survey data.
The main difference between the ground and space configurations
is that from space, the statistics are significantly improved, and so
the detection contours are significantly narrower.

\subsection{The effect of assigning redshifts from an assumed distribution}

To quantify the uncertainty arising from not knowing the redshifts
for background sources, we performed the analysis after assigning
redshifts drawn from a distribution. The 25000 sources are put at a mean redshift
$z_s$ and images are simulated. When constructing the simulated catalogue,
the background objects are assigned a mean redshift of $z_s+\delta_z$.
This catalogue is then input into the maximum likelihood code. Since the
strength of the shear is proportional to the distance between the
sources and the lenses, under estimating the source redshifts leads to
an overestimate of the lens masses and tends to shift the
confidence contours toward higher values for the velocity dispersion. 
For the same reason, systematically overestimating the
source redshifts leads to an underestimate of the galaxy masses.  In
any case, we find that a redshift uncertainty of $\pm 0.2$ does not dramatically
modify the conclusions.  This is the typically the precision we can get
with photometric redshift estimation which is encouraging for future surveys.
This results is coherent with a recent study by Kleinheinrich et al., 2004 
based on the galaxy-galaxy lensing results from the COMBO-17 survey.
They found that it is of great importance to know the redshifts of individual
lens galaxies in order to constrain the properties of their dark matter
halos, but that the knowledge of individual source redshifts improves the
measurements only very little over use of statistical source redshift
distribution.

To be more quantitative, let us consider the lensing equation and express
it for a constant deflection angle. For a PIEMD profile with a given
$\sigma_0$, we have:
\begin{equation}
\sigma_0^2 \frac{D_{LS}}{D_S} = \mathrm{constant}; \quad {\rm and} \quad 
\frac{D_{LS}}{D_S}=E(z_l,z_s) 
\end{equation}
as introduced in \cite{ggolse}. Therefore the equation can be
rewritten as:
\begin{equation}
\sigma_0^2 E(z_L,z_S) = \mathrm{constant}
\end{equation}
The lenses are kept at a redshift of 0.2, and the mean redshift of the sources
is changed by $\delta z=\pm$ 20 \%. We then evaluate the 
corresponding $\delta \sigma_0$ error introduced in the retrieval of
the central velocity dispersion in the likelihood analysis.
Table \ref{deltaz} gives the results: when we put an error of $\delta z_S$ (\%), this give a variation of $E(z_L,z_S)$ equal to $\delta E$, and the 
corresponding variation on $\sigma_0$ is of order $\delta \sigma_0$. 
This range of values is given by the projection of the 3$\sigma$ contours
along the $\sigma_0$ axis.
We can see that the variation in the detection range is in agreement with 
the calculations made.
\begin{table}
\begin{center}
\begin{tabular}{|c|c|c|c|}
\hline
\hline
$\delta z_S$ (\%)  &  $\delta E$ & $\delta \sigma_0$ &  $\sigma_0$ \\
\hline
0 & 0 & 0 & [174-196] \\
\hline
20  & 0.05 & -7 & [170-190] \\
\hline
-20  & -0.05 & 7 & [180-208] \\
\hline
\hline
\end{tabular}
\caption{Influence of an uncertainty in the mean redshift of the sources: 
an error of $\delta z_S$ correspond to a variation of $E(z_L,z_S)$ equal to $\delta E$,
which correspond to a variation on $\sigma_0$ of order $\delta \sigma_0$. 
We see from this table that this estimated variation on $\sigma_0$ is coherent
with the variation in the detection range as derived from the maximum likelihood analysis.}
\label{deltaz}
\end{center}
\end{table}

\subsection{Influence of the uncertainty in the cluster modeling}

When working with the real data, we will have to put in by hand the
description of the cluster. The reliability of the results will depend
on the accuracy with which we describe cluster. 
In order to study the influence of the uncertainty of the cluster
profile, the cluster component is described by a PIEMD profile with a
velocity dispersion of $\sigma_{cluster}=1000\,\mathrm{kms}^{-1}$.  When
constructing the simulated catalogue, the cluster component is assigned
a velocity dispersion of $\sigma_{cluster}+\delta_{\sigma}$. The
likelihood is then computed for this case.

\section{Discussion and Conclusions}

\subsection{Constraints obtained on mass profiles}

For the PIEMD and NFW profiles, we found that we are able to retrieve
the characteristic halo parameters with extremely good accuracy, for
every configuration. In fact, interestingly enough, the dotted lines
in Figs. 9, 10, and 11 show us that the aperture mass is retrieved
very accurately. This immediately suggests the re-parameterisation of
models considered here: rather than fitting in the ($\sigma_0,r$)
plane, we can fit the deformations directly in the ($M_{aper},
R_{aper}$) plane. This formulation is explored in the next section.

For the PL profile, we find that we can put some constraints
on $\alpha$, the slope of the density profile, but not on the 
velocity dispersion, since the likelihood function does not always 
converge along that direction. So we can use this profile to 
estimate the slope of dark matter halos, without trusting the constraints 
we get on the velocity dispersion for the profile.

\subsection{Influence of $R_{max}$}

\begin{figure}
\begin{center}
\includegraphics[height=4cm,width=4cm]{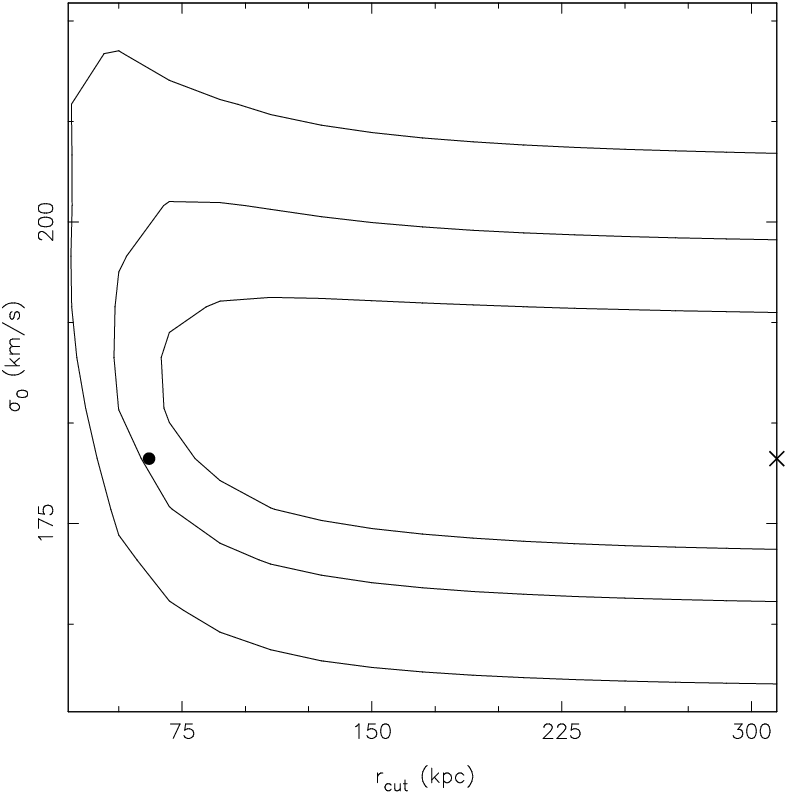}
\includegraphics[height=4cm,width=4cm]{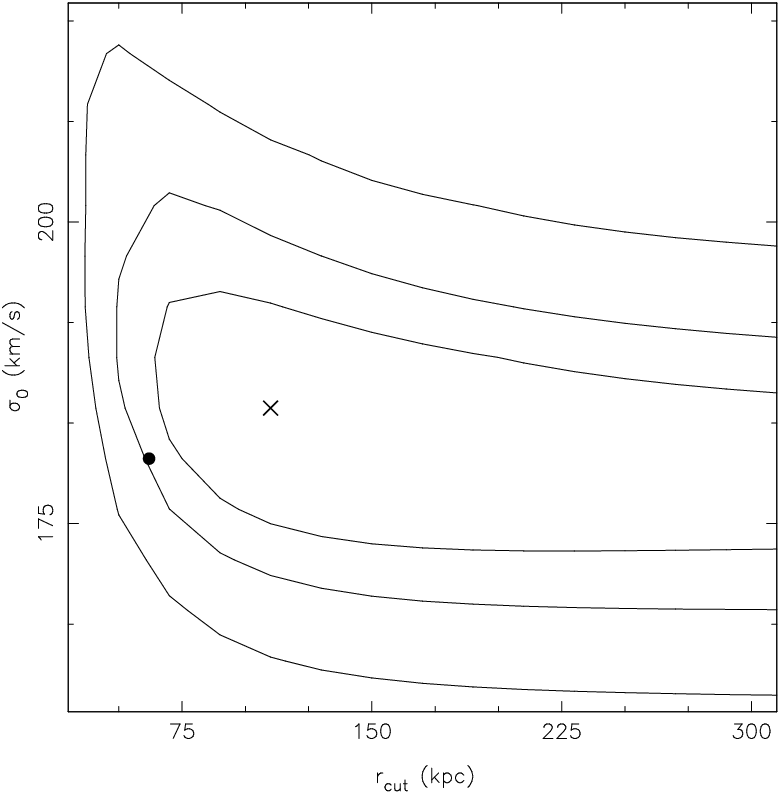}\\
\includegraphics[height=4cm,width=4cm]{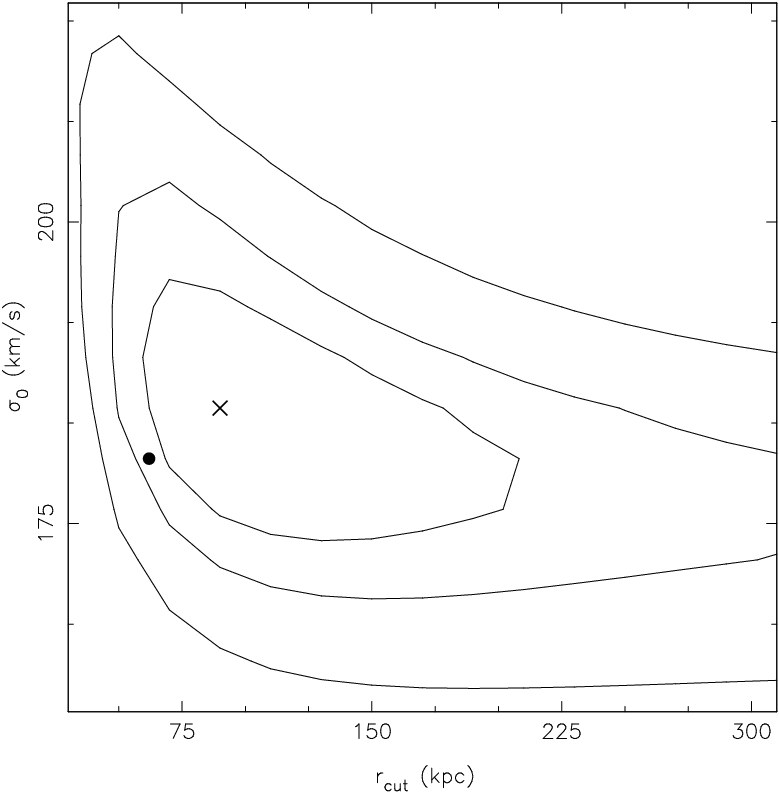}
\includegraphics[height=4cm,width=4cm]{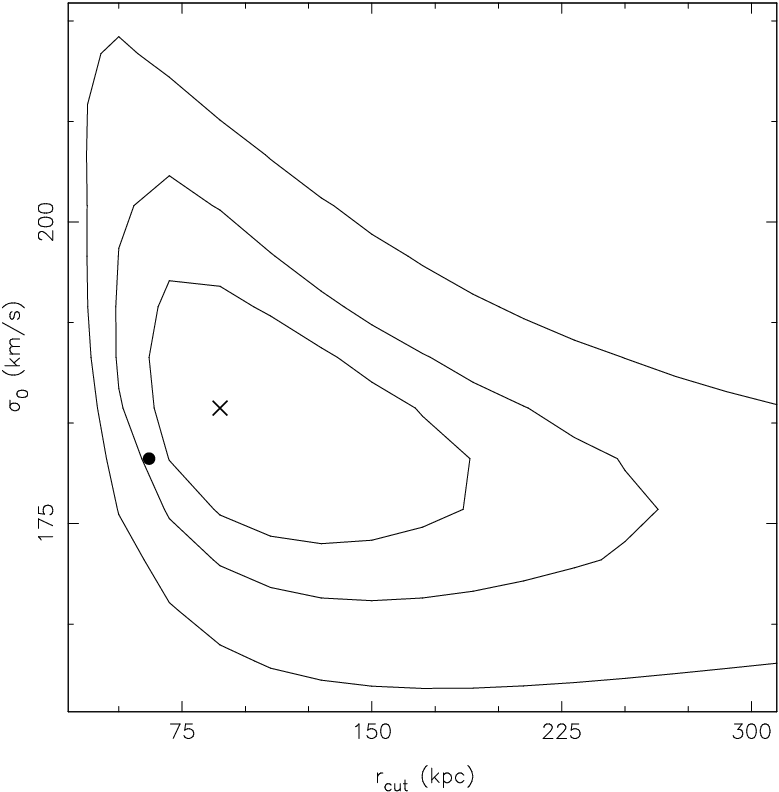}
\end{center}
\caption{\label{fig5} The PIEMD profile, in a field
  configuration: In the panels from left to right, the value of
  $R_{max}$ increases from 30", 60", 90" to 150". When $R_{max}>$ 100",
  the contours coverge along the $r_{cut}$ axis. Note that the
  convention throughout this work is that the dot marks the value of the input
  parameters and the cross marks the retrieved output values.}
\end{figure}

When working with real data, the results we get depend on the value
chosen for $R_{max}$.  When we take a low value for $R_{max}$, the
shear for the image (i) is calculated with fewer lenses, and we find
that the contours do not close in $r_{cut}$. When the value of this
parameter is increased , the contours converge and close. On the other
hand, picking a high value for $R_{max}$ introduces some noise in
calculating the shear, and can dilute the galaxy-galaxy lensing signal
significantly: lenses that do not \emph{effectively} participate in
the lensing of an image if utilised in the calculation become a source
of noise. Fig.~\ref{fig5} illustrates precisely this situation: for a
given value of $R_{max}$, we obtain a good estimate of $\sigma_0$, but
the robustness of the constraint on $r_{cut}$ is directly related to
the value of $R_{max}$.  Others authors have reported that their
results are sensitive to the value of this parameter (\emph{e.g.}
Kleinheinrich, PhD thesis).

\begin{figure}
\begin{center}
\includegraphics[height=7cm,width=7cm]{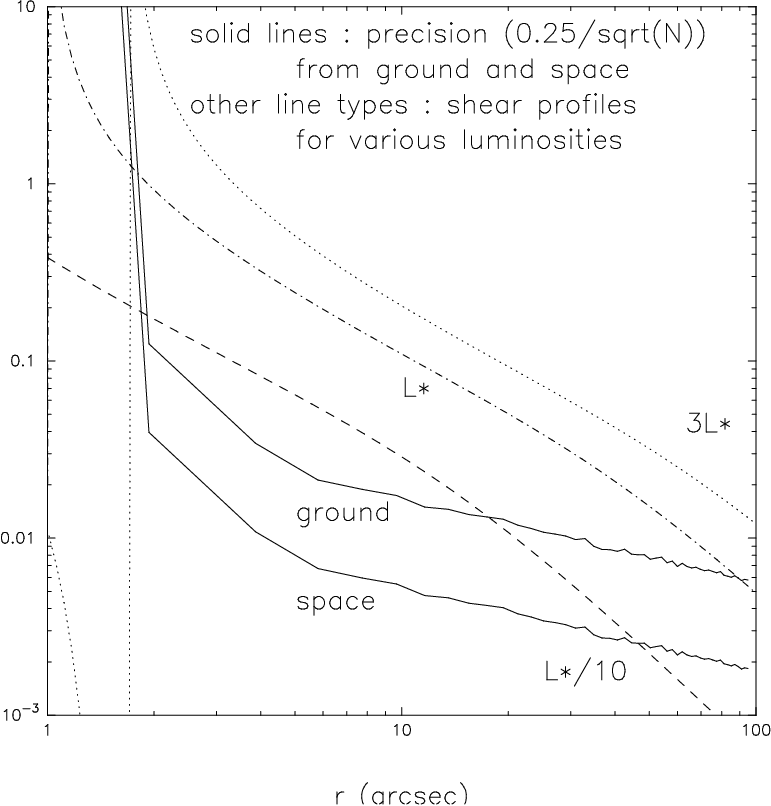}
\end{center}
\caption{\label{fig6} Comparing the signal to noise for an
  estimate of the optimal choice for $R_{max}$. The solid lines
  correspond to the characteristic ellipticity (due to the width 
of the ellipticity distribution), $\frac{0.25}{\sqrt{N}}$ as a 
function of radius, from ground and space. The other line types 
(dot, dashed and dot-dashed) correspond to the signal - the reduced 
shear as a function of radius for typical luminosities.}
\end{figure}

The choice of $R_{max}$ is therefore important. To get the order of magnitude
of this parameter, we compare the characteristic noise in the problem,
i.e. $\frac{0.25}{\sqrt{N}}$, to the signal 
we are sensitive to i.e. the reduced shear - the factor of 0.25 is the
width of the intrinsic ellipticity distribution and N is the number 
of background objects at a distance $r$
from a lensing galaxy. This noise has been estimated by analysing data from the 
ground-based CFH12k observations of the cluster A1763 at a
redshift of $z = 0.22$. From space, we expect the number of background
objects to be about 6/7 times higher. Fig.~\ref{fig6} shows that a 
value of about 100" can be used for $R_{max}$.

On the other hand, the choice of the parameter $R_{min}$ does not
influence the results, so we fix this parameter to be of the order of a few kpc. 

\section{Re-parameterization of the problem}

Thus far, we have performed the likelihood analysis to optimize the values 
of two parameters $\sigma_0$ and $r_{cut}$ (or $r_{s}$). A different
set of parameters can be chosen for maximizing the likelihood, for instance
$M_{aper}$ and $R_{aper}$, which we explore below.

We have:
\begin{equation}
\label{ML_S_R}
\mathcal{L}=\mathcal{L}(\sigma_0, r_{cut})\quad \& \\ 
M_{aper}=M_{aper}(R_{aper},\sigma_0,r_{cut})
\end{equation}
so we can write:
\begin{equation}
\sigma_0=\sigma_0(M_{aper},R_{aper},r_{cut})
\end{equation}
the likelihood function then becomes:
\begin{equation}
\mathcal{L}(M_{aper},R_{aper},r_{cut})
\end{equation}
and by summing over $r_{cut}$, we get:
\begin{equation}
\mathcal{L'}=\sum_{r_{cut}} \mathcal{L}(M_{aper},R_{aper},r_{cut})
\end{equation}

\begin{equation}
\mathcal{L'}=\mathcal{L'}(M_{aper},R_{aper})
\end{equation}

\begin{figure}
\begin{center}
\includegraphics[height=7cm,width=7cm]{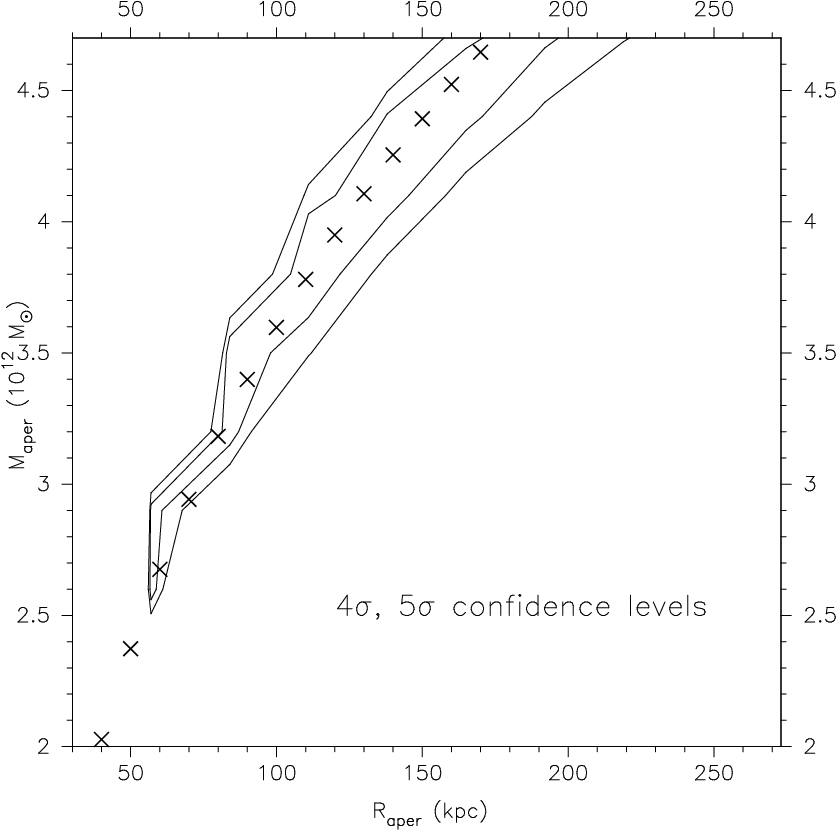}
\includegraphics[height=7cm,width=7cm]{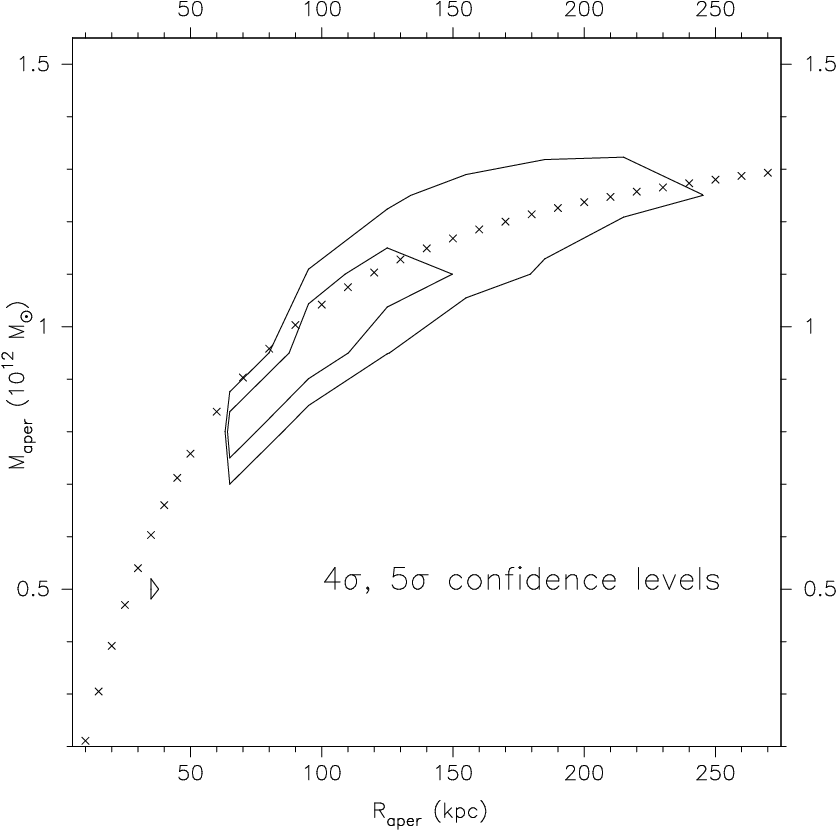}
\caption{\label{fig7}
The likelihood $\mathcal{L'}(M_{aper},R_{aper})$ for the NFW (upper panel) and the
  PIEMD profile (lower panel). The crossed line represents the
  $M_{aper}(R_{aper})$ contour obtained with the input parameters used
  to describe the foreground lenses in the {\bf cluster} configuration.} 
\end{center}
\end{figure}

\begin{figure}
\begin{center}
\includegraphics[height=7cm,width=7cm]{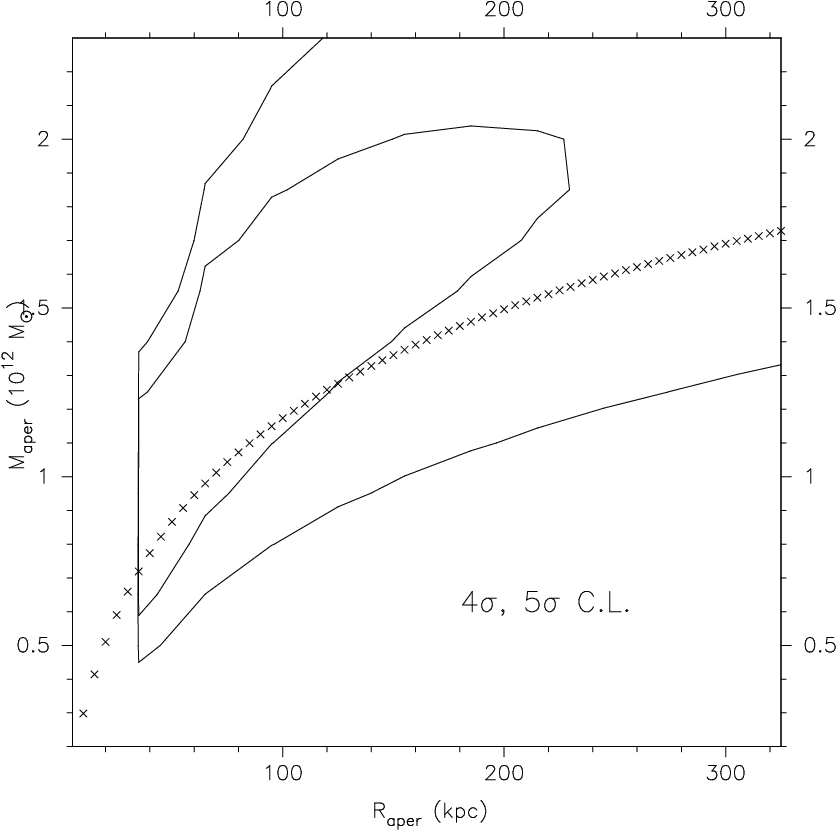}
\includegraphics[height=7cm,width=7cm]{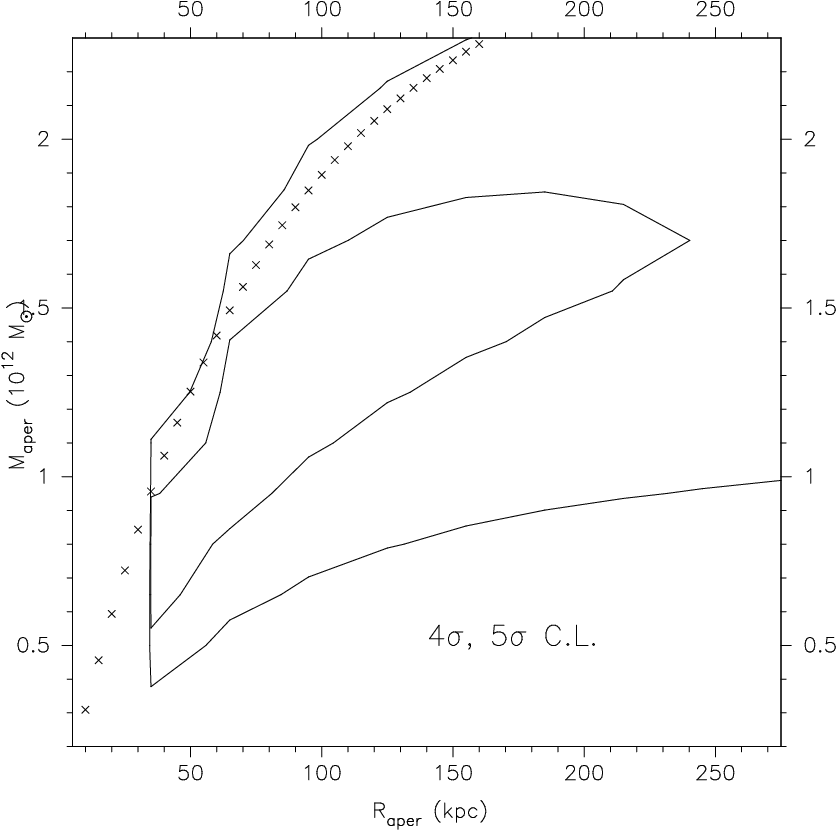}
\caption{\label{fig8}
  The likelihood $\mathcal{L'}(M_{aper},R_{aper})$ for the NFW (upper panel) and the
  PIEMD profile (lower panel). The crossed line represents the
  $M_{aper}(R_{aper})$ contour obtained with the input parameters used
  to describe the foreground lenses in the {\bf field} configuration.} 
\end{center}
\end{figure}

\begin{figure*}
\begin{center}
\includegraphics[width=5.5cm,height=5.5cm]{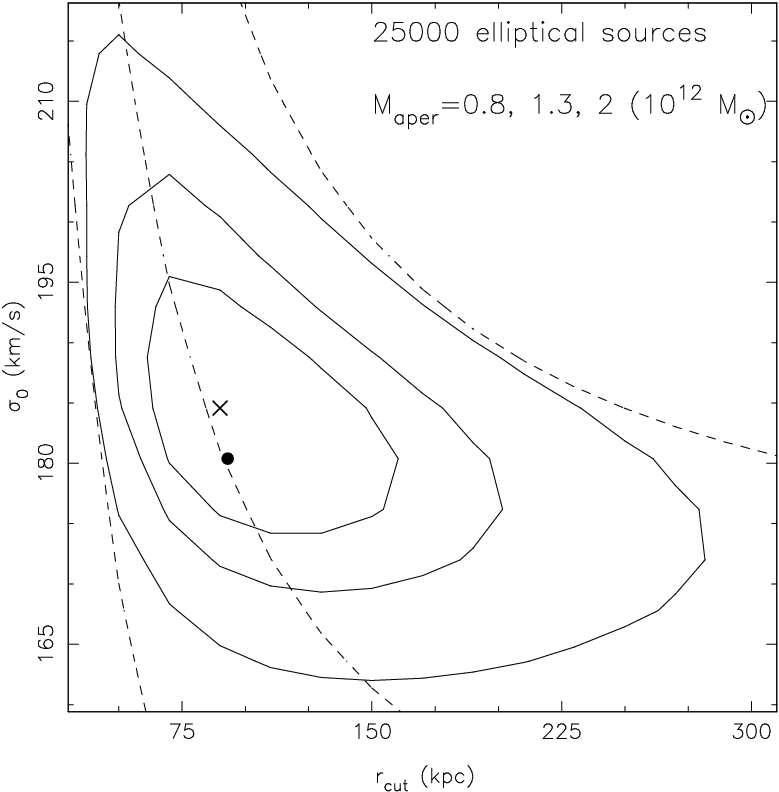}
\includegraphics[width=5.5cm,height=5.5cm]{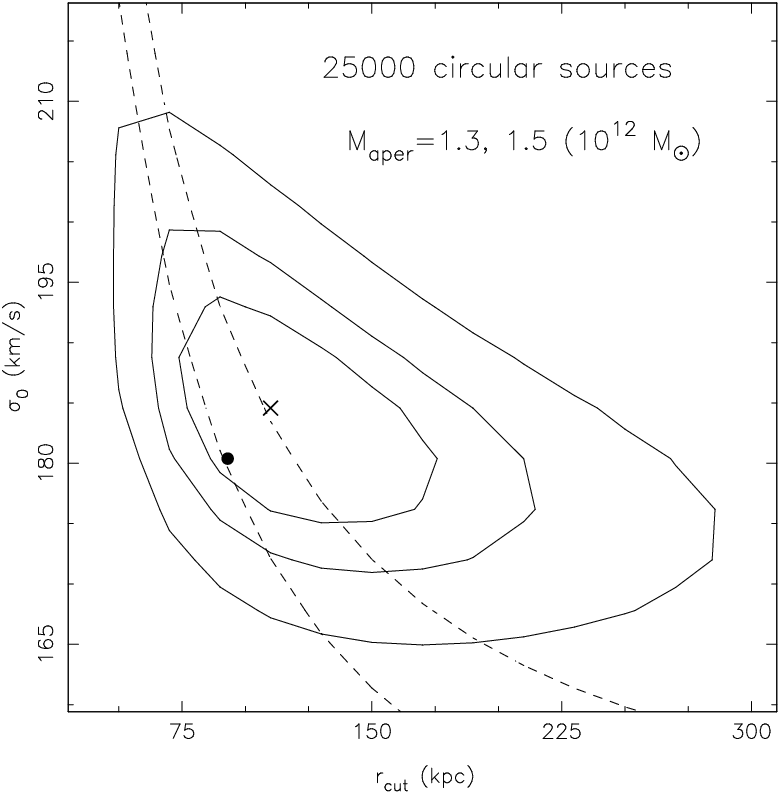}
\includegraphics[width=5.5cm,height=5.5cm]{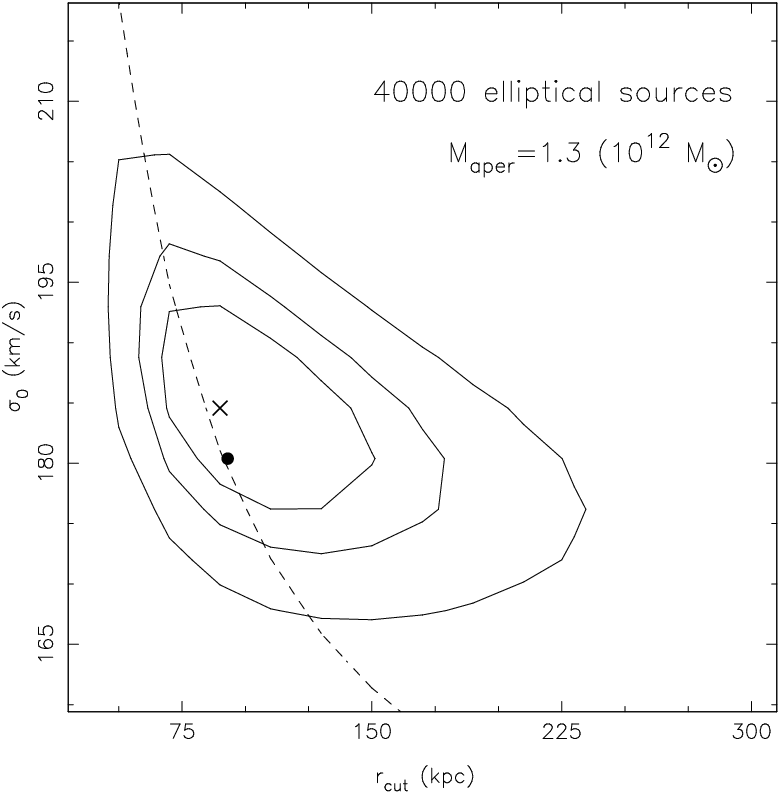}\\
\includegraphics[width=5.5cm,height=5.5cm]{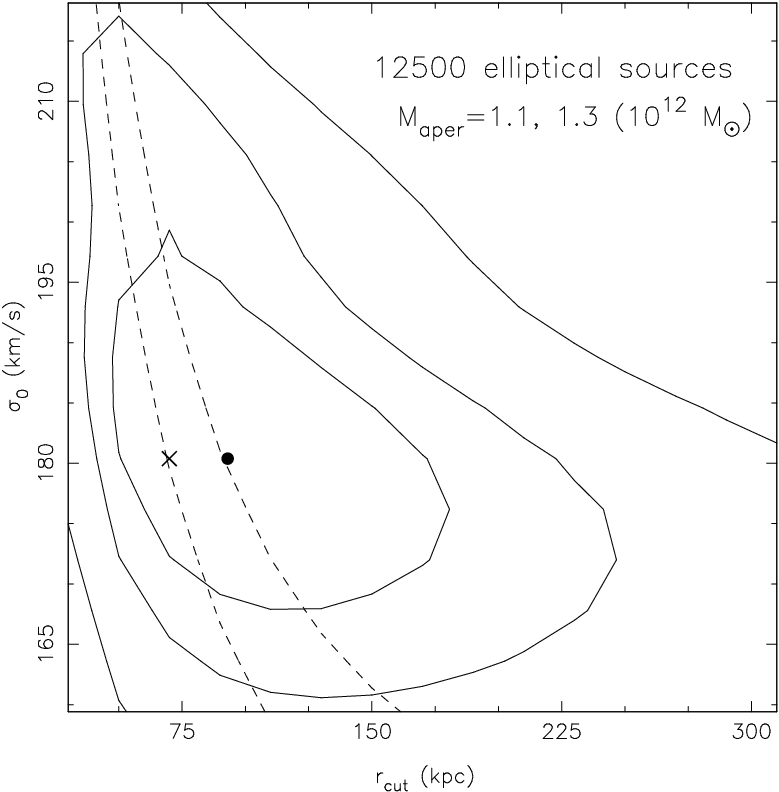}
\includegraphics[width=5.5cm,height=5.5cm]{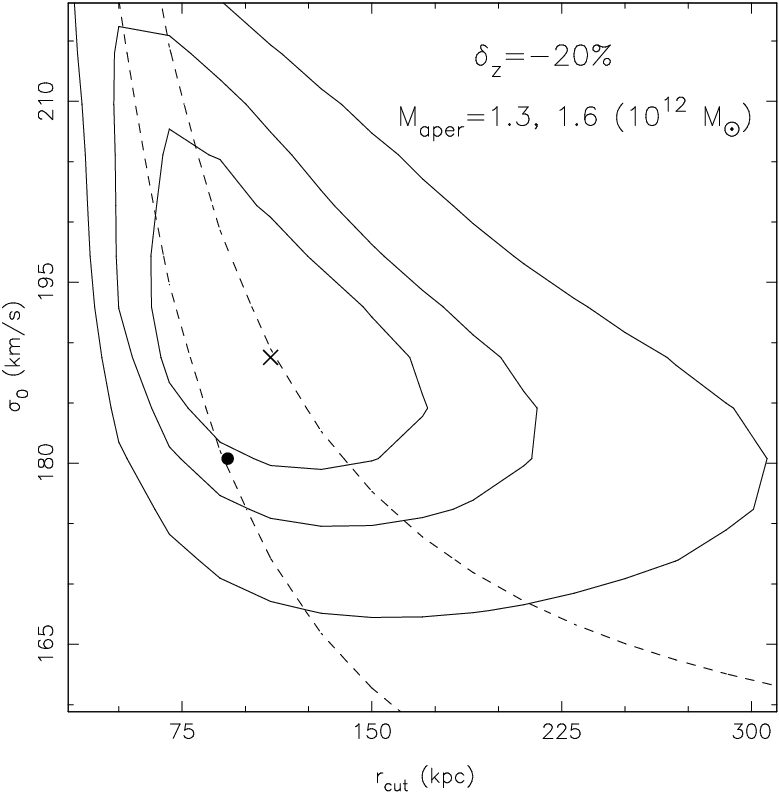}
\includegraphics[width=5.5cm,height=5.5cm]{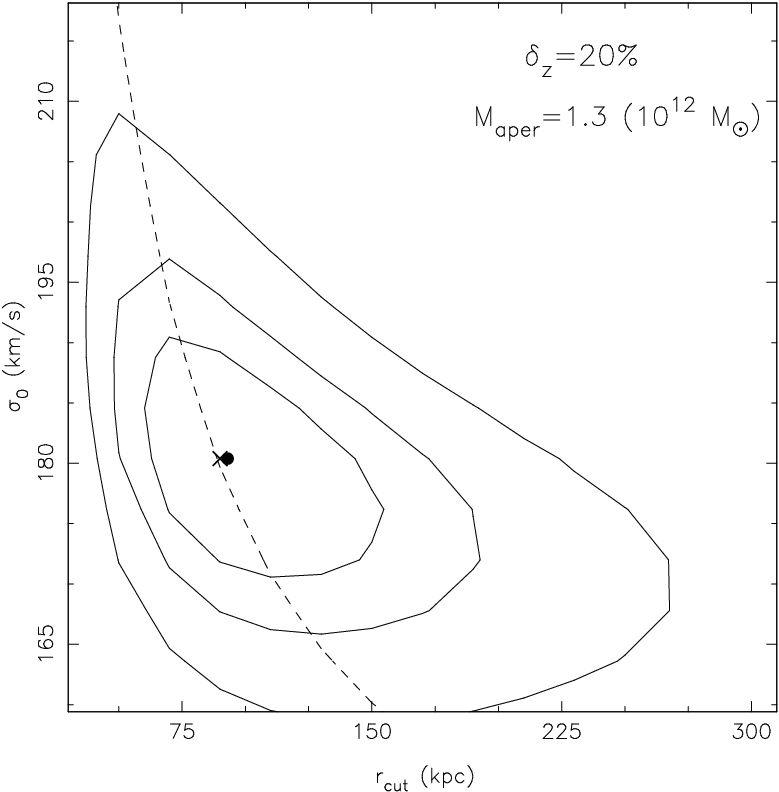}\\
\includegraphics[width=5.5cm,height=5.5cm]{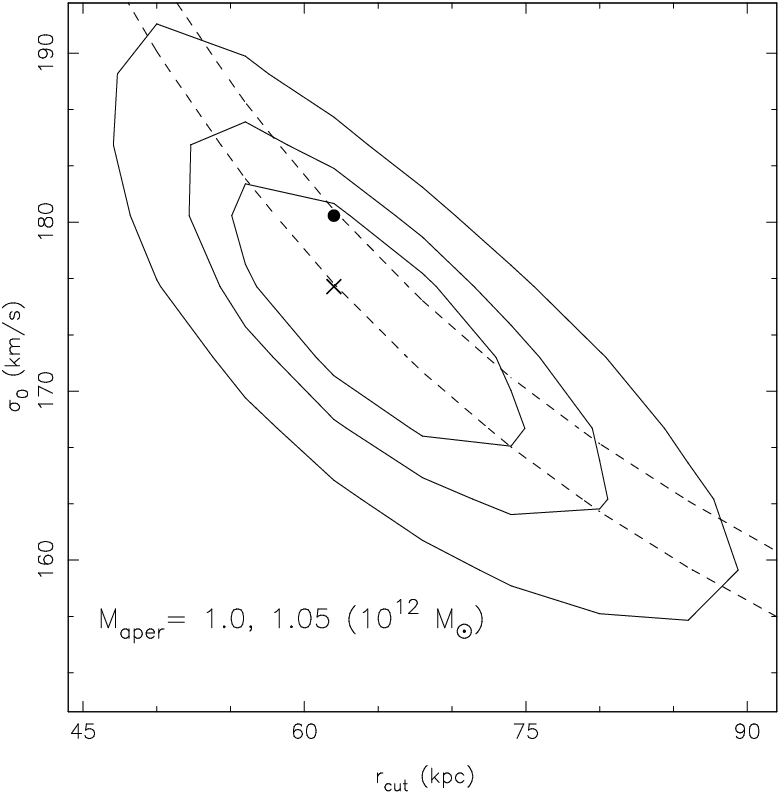}
\includegraphics[width=5.5cm,height=5.5cm]{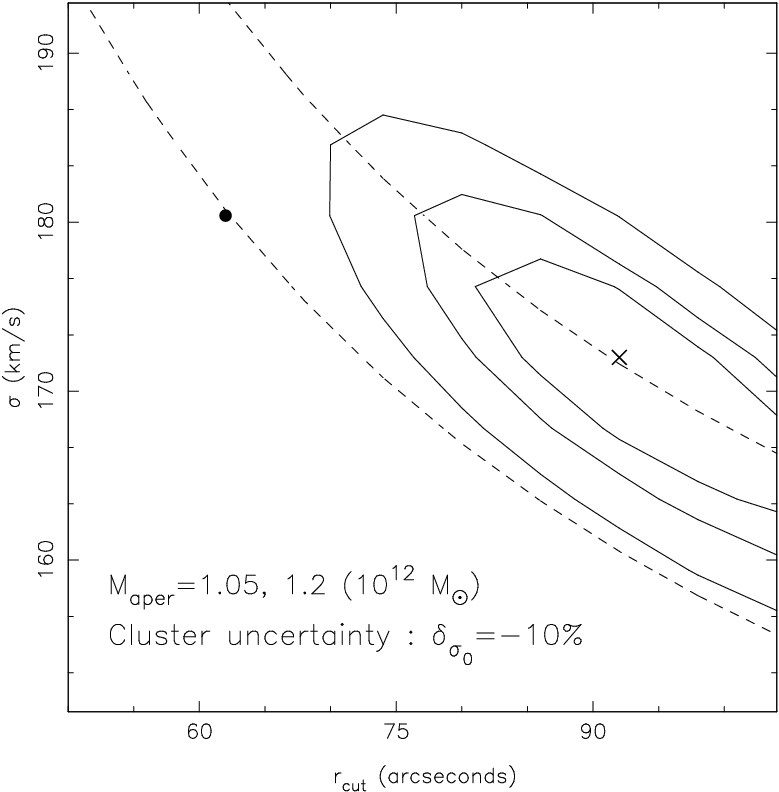}
\includegraphics[width=5.5cm,height=5.5cm]{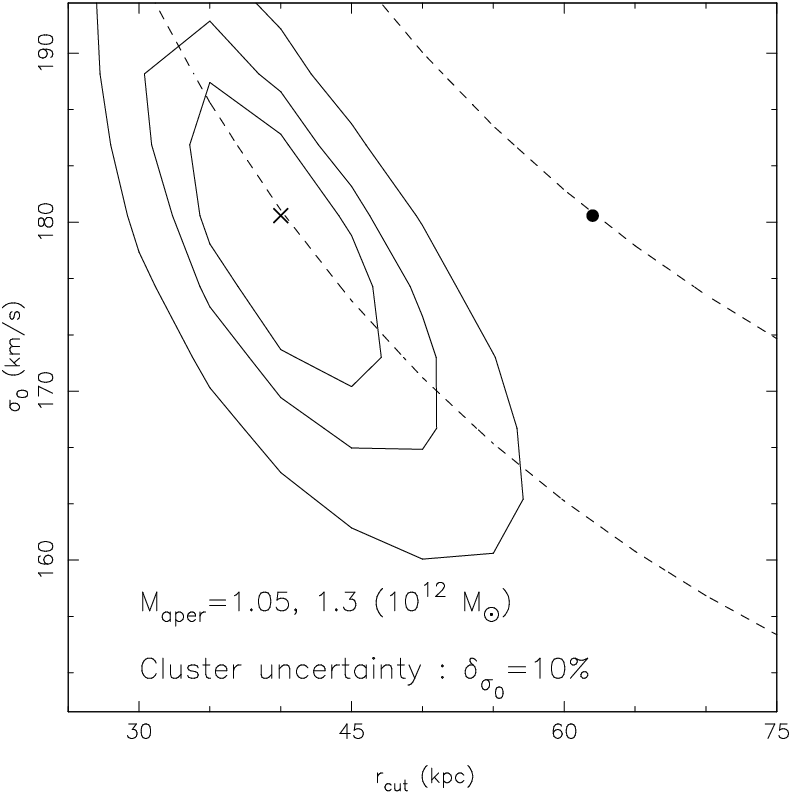}
\end{center}
\caption{\label{fig9} Results for the PIEMD
  profile. 
The top left plot (number 1) is the reference field situation with 25,000
  elliptical background sources in a field 
of 26$\times$26 arcmin$^2$, corresponding to a density of background sources equal to about 35 galaxies per arcmin$^2$.
Then the following plots (2, 3, 4) show the results obtained with 25,000 \emph{circular} sources,
then with 40,000 elliptical sources mimicking typical space observations with roughly 60 galaxies per arcmin$^2$, then with
  12,500 elliptical sources, corresponding to the ground based configuration with a galaxy number density of 17 galaxies per arcmin$^2$. 
The plots number 5 and 6 show the effect of introducing an uncertainty in the
  mean redshift of the source population: this uncertainty is equal  
to -20\% (left) and +20\% (right).
The last row show the results from the cluster configuration: the
  reference cluster configuration (plot 7), then
  configurations where an uncertainty on the cluster
  modeling is introduced: knowing the central velocity dispersion of the cluster to
within $\pm 10\%$. The contours in this figure represent the 3$\sigma$,4$\sigma$,5$\sigma$
confidence levels, and along the dotted lines in every panel, the mass
  within a projected radius $R_{aper}$ of 100 kpc is kept constant at
  the value indicated on the plot. Note
  that the dot indicates input values and the cross the retrieved output.} 
\end{figure*}

\begin{figure*}
\begin{center}
\includegraphics[width=5.5cm,height=5.5cm]{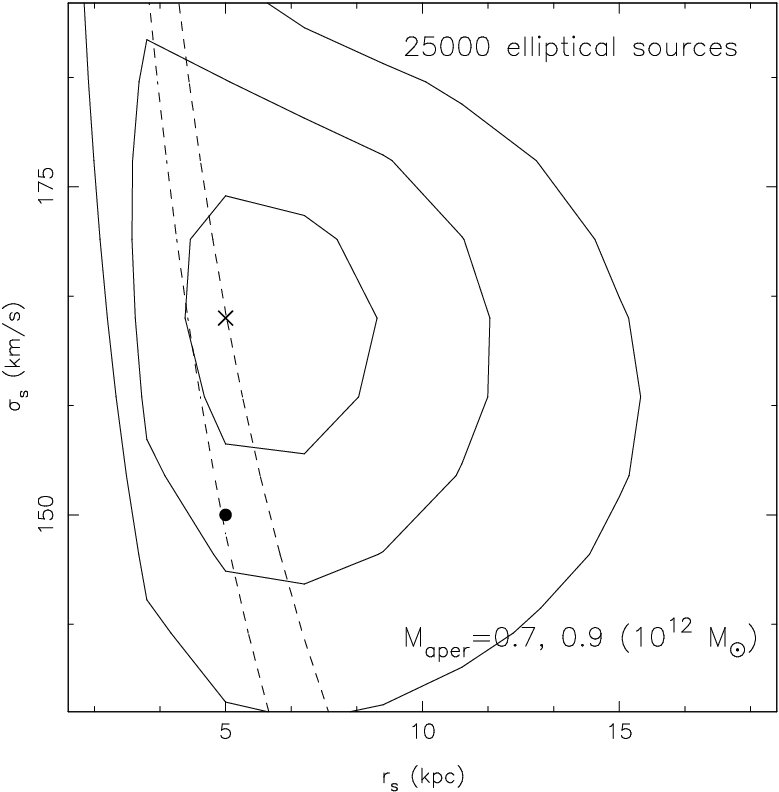}
\includegraphics[width=5.5cm,height=5.5cm]{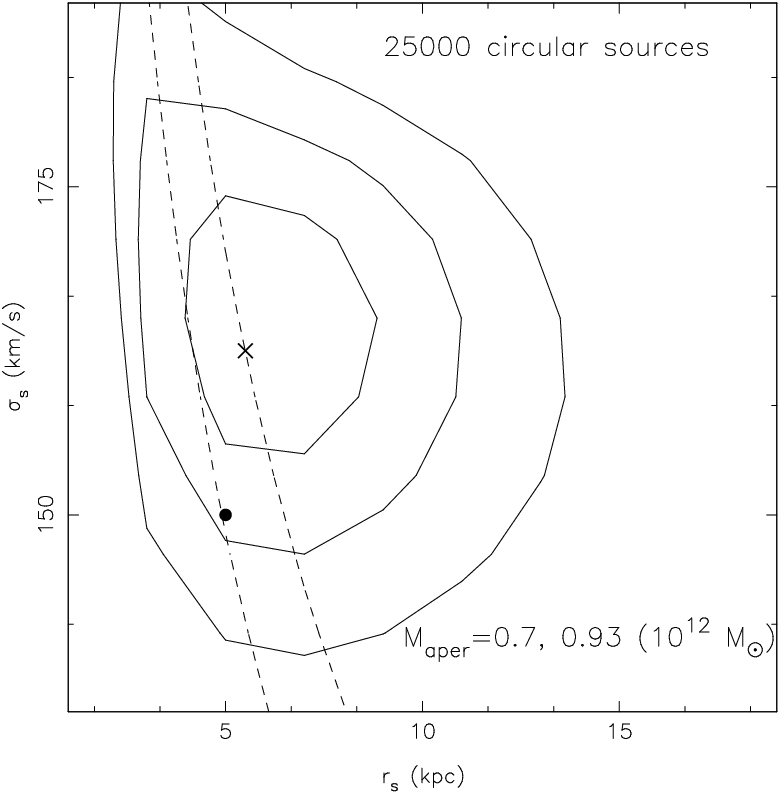}
\includegraphics[width=5.5cm,height=5.5cm]{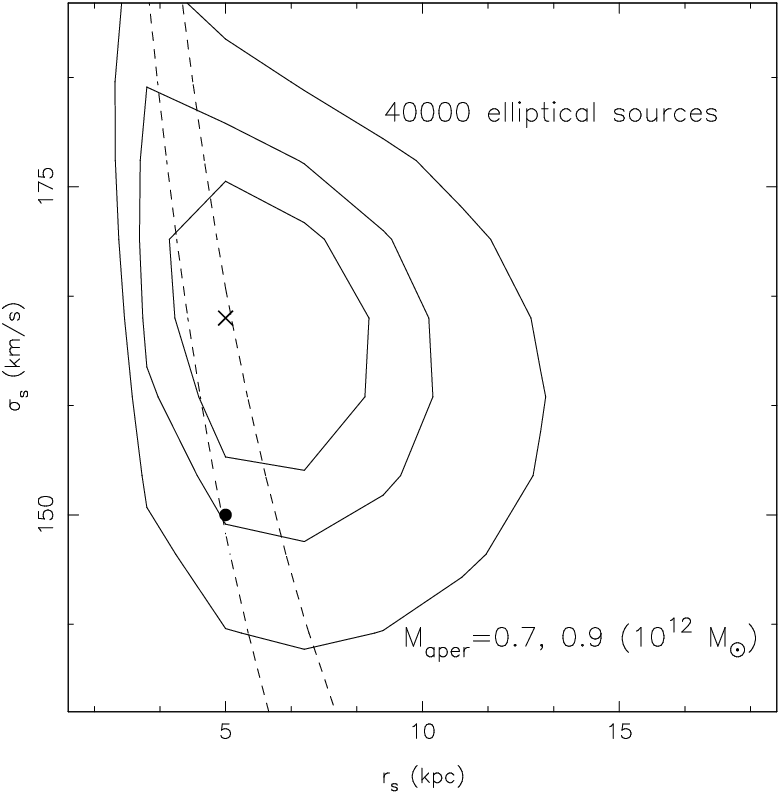}
\includegraphics[width=5.5cm,height=5.5cm]{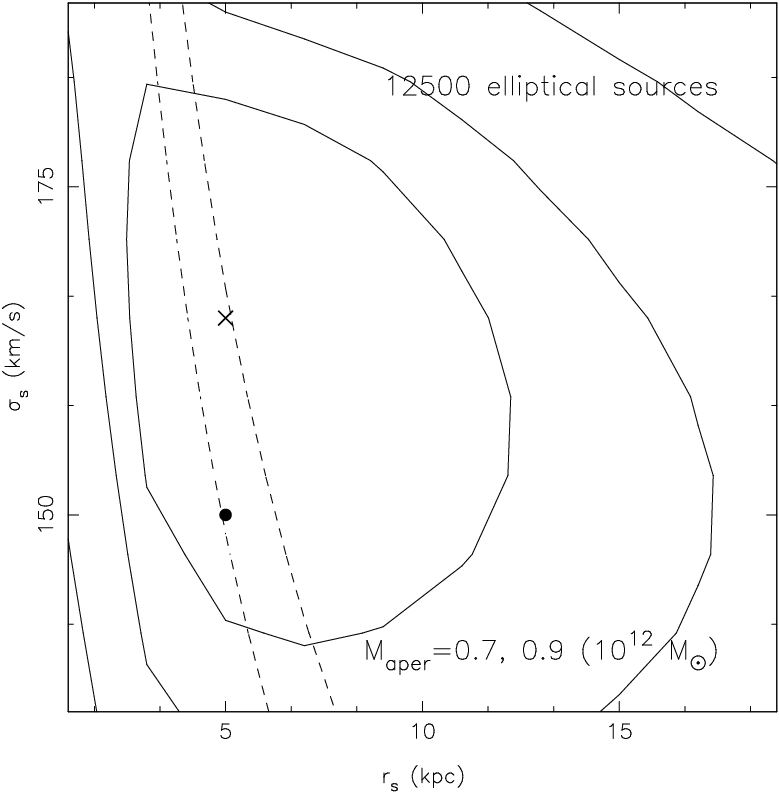}
\includegraphics[width=5.5cm,height=5.5cm]{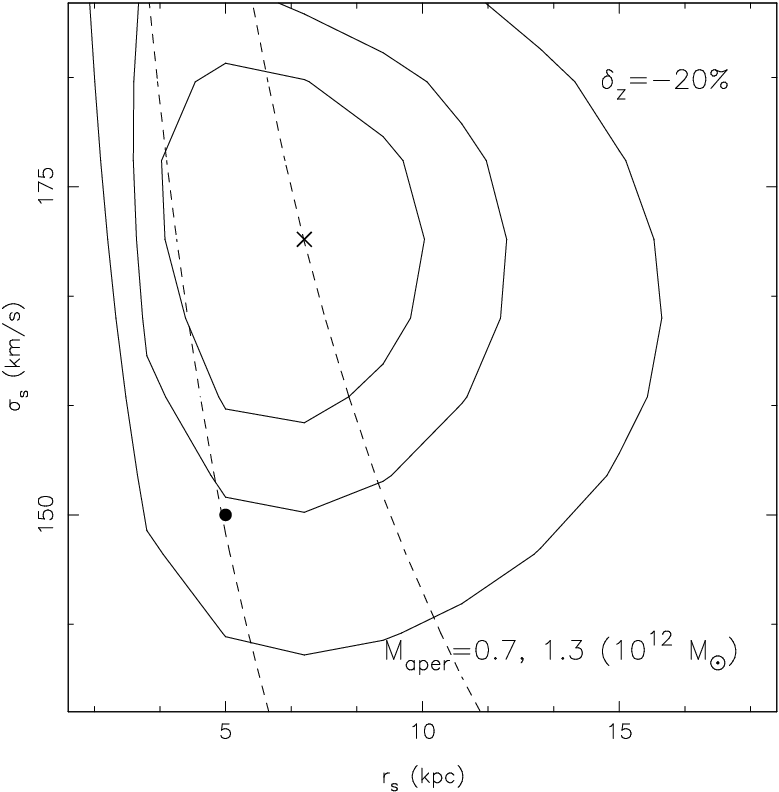}
\includegraphics[width=5.5cm,height=5.5cm]{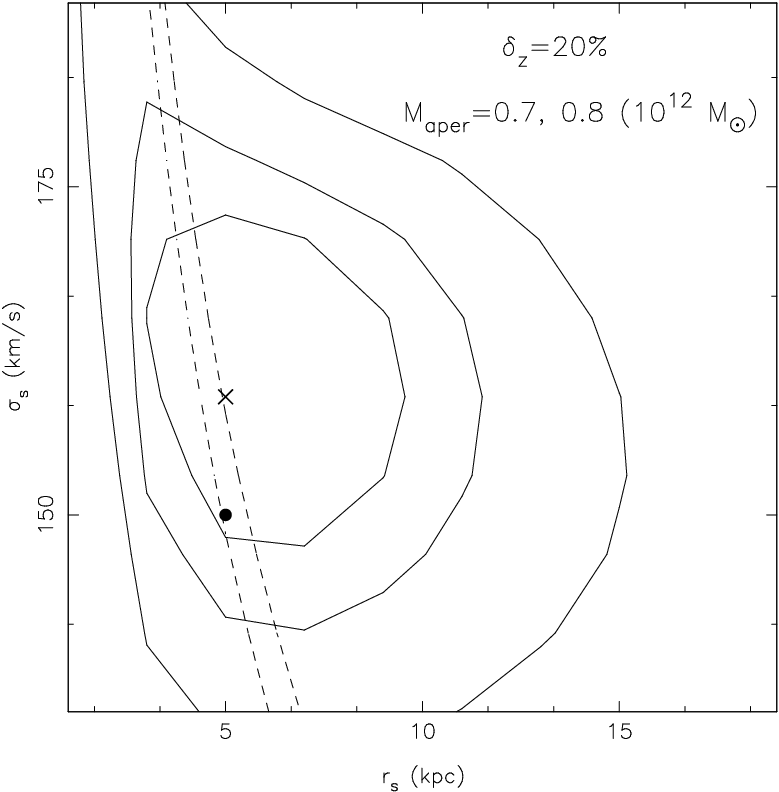}
\includegraphics[width=5.5cm,height=5.5cm]{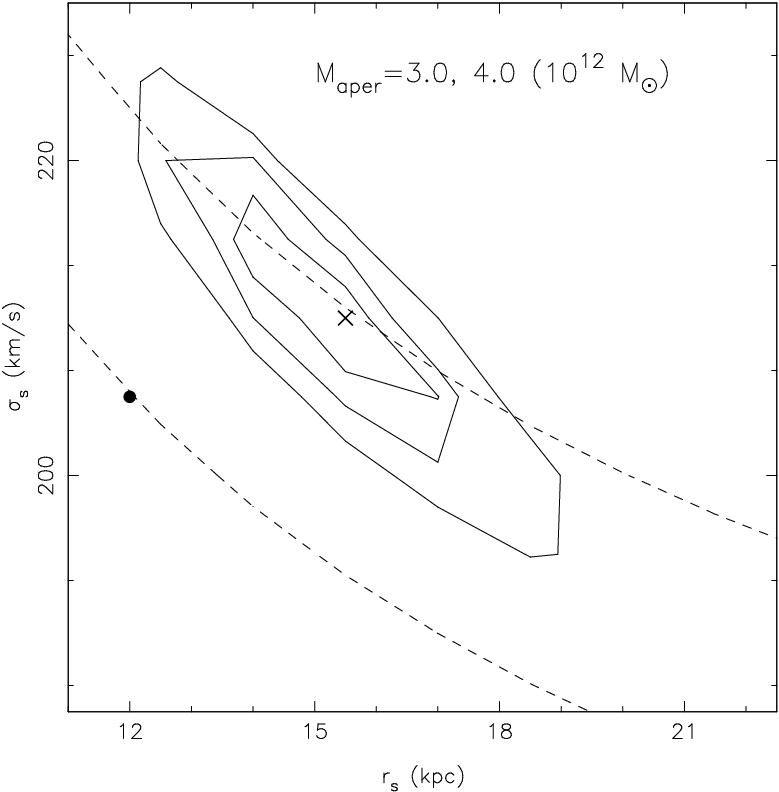}
\includegraphics[width=5.5cm,height=5.5cm]{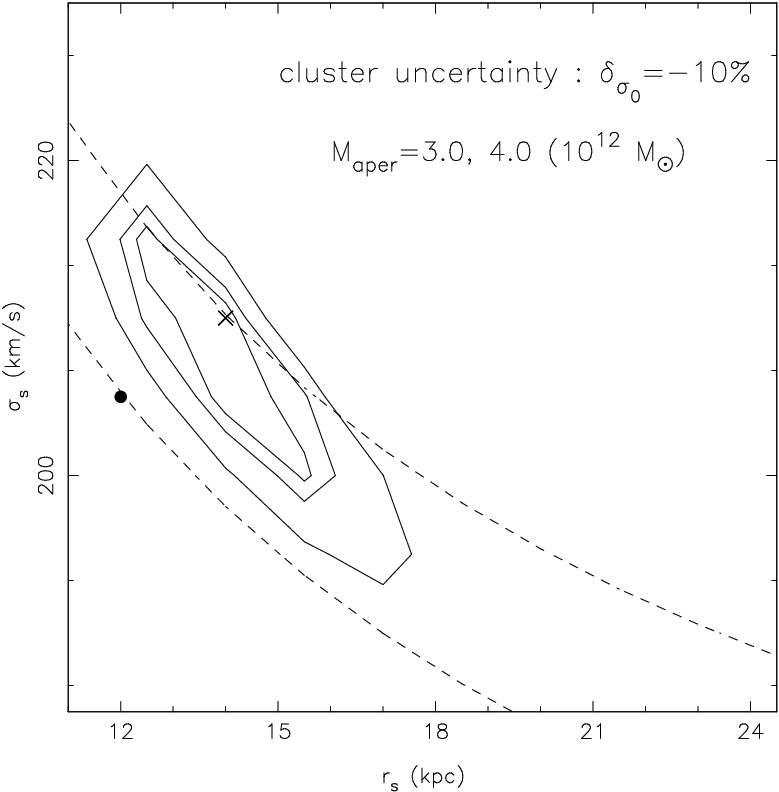}
\includegraphics[width=5.5cm,height=5.5cm]{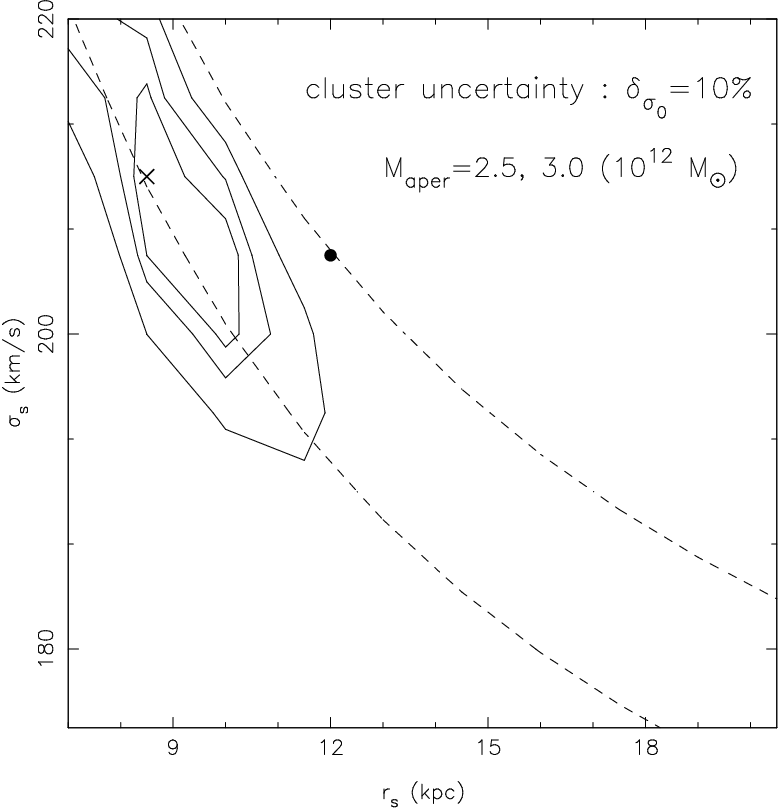}
\end{center}
\caption{\label{fig10} Results from the NFW profile:
  the top left plot (number 1) is the reference field situation with 25,000 elliptical background sources in a field of 26$\times$26 arcmin$^2$,
corresponding to a density of background sources equal to about 35 galaxies per arcmin$^2$.
The following plots (2, 3, 4)  show the results obtained with 25,000 \emph{circular} sources,
then with 40,000 elliptical sources, corresponding to data obtained
  from space with 60 galaxies per arcmin$^2$, then with 12,500 elliptical
  sources, corresponding to the ground-based data of about 17 galaxies per arcmin$^2$.
The plots number 5 and 6  show the effect of introducing an uncertainty in the
  mean redshift of the source population: this uncertainty is equal
to -20\% (left) and +20\% (right). 
The last row show the results from the cluster configuration:
 the reference cluster configuration (plot 7), the effect of
  introducing an uncertainty of $\pm10\%$ in the central velocity
  dispersion of the cluster model used. Contours in these figures
  represent the 3$\sigma$, 4$\sigma$, and 5$\sigma$ confidence levels,
  and along the dotted lines in each panel, the mass within a
  projected radius $R_{aper}$ of 100 kpc is kept constant,
with $M_{\rm aper}$ at the quoted value indicated on the plot. Note
  that the dot indicates input values and the cross the retrieved output.}
\end{figure*}

\begin{figure*}
\begin{center}
\includegraphics[width=5.5cm,height=5.5cm]{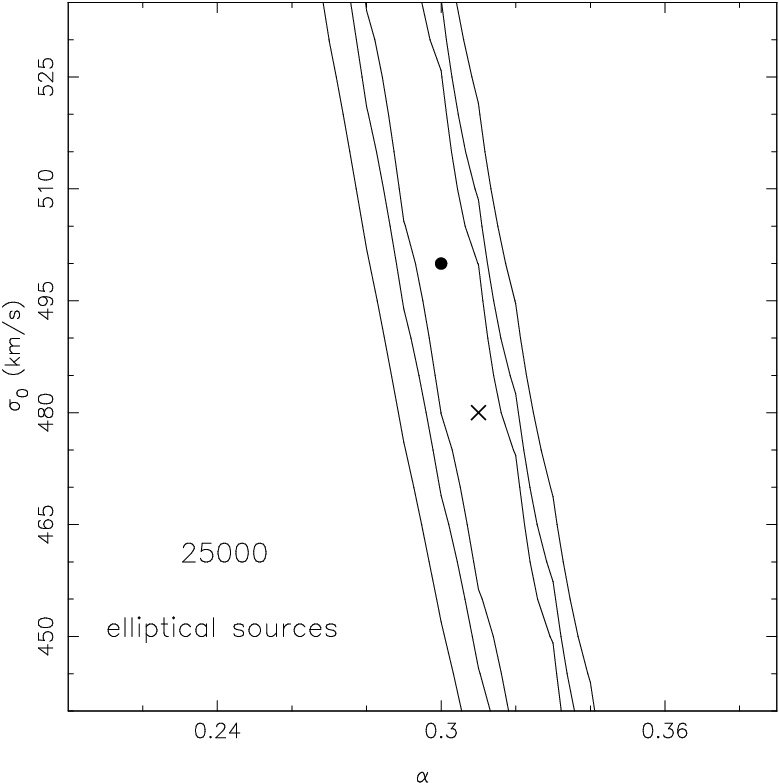}
\includegraphics[width=5.5cm,height=5.5cm]{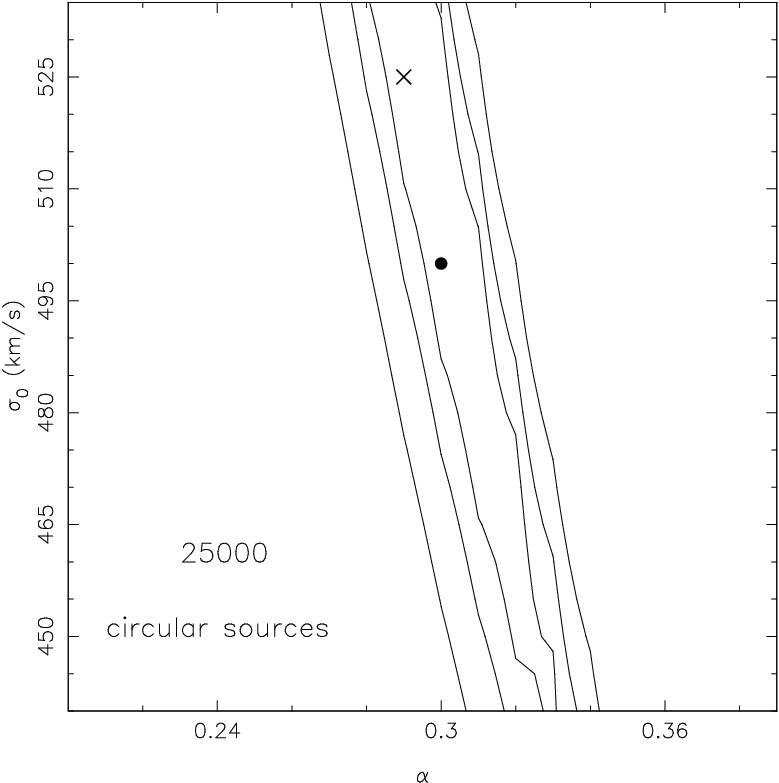}
\includegraphics[width=5.5cm,height=5.5cm]{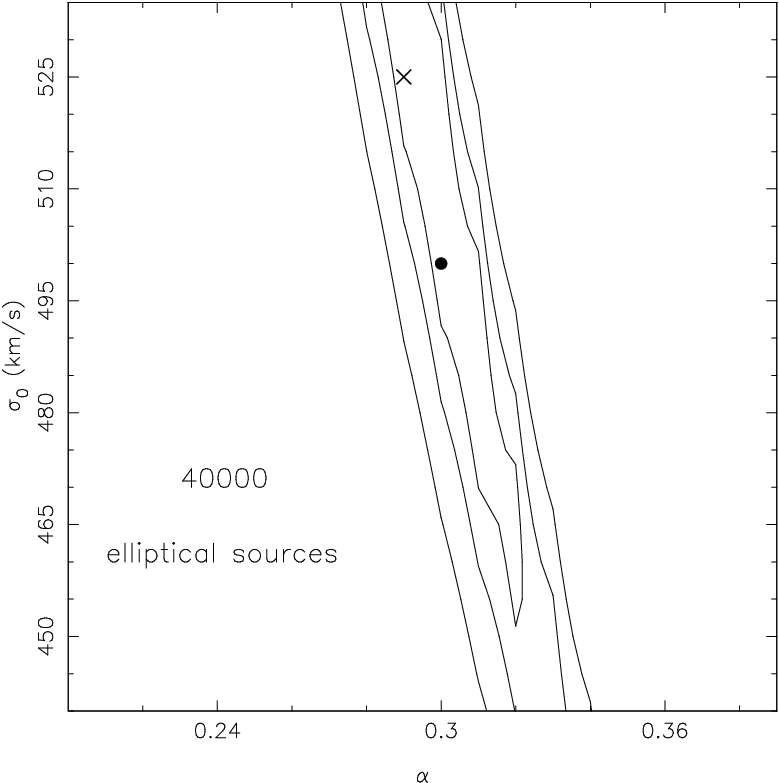}
\includegraphics[width=5.5cm,height=5.5cm]{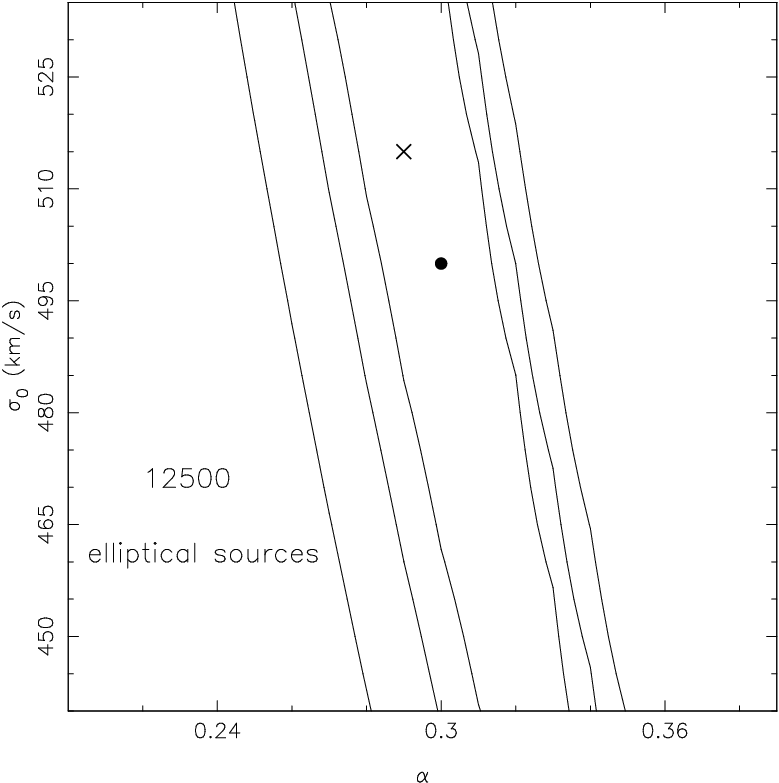}
\includegraphics[width=5.5cm,height=5.5cm]{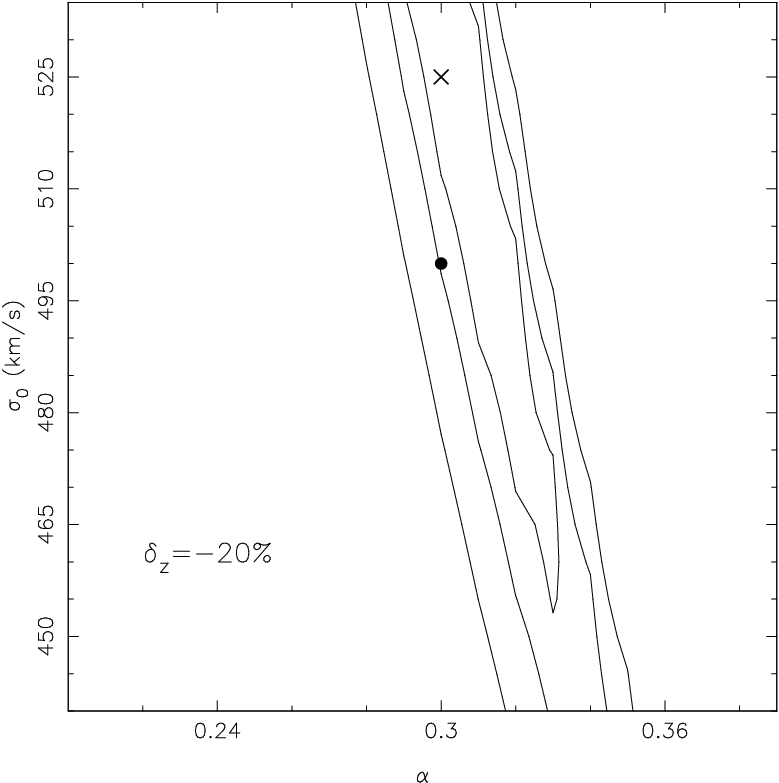}
\includegraphics[width=5.5cm,height=5.5cm]{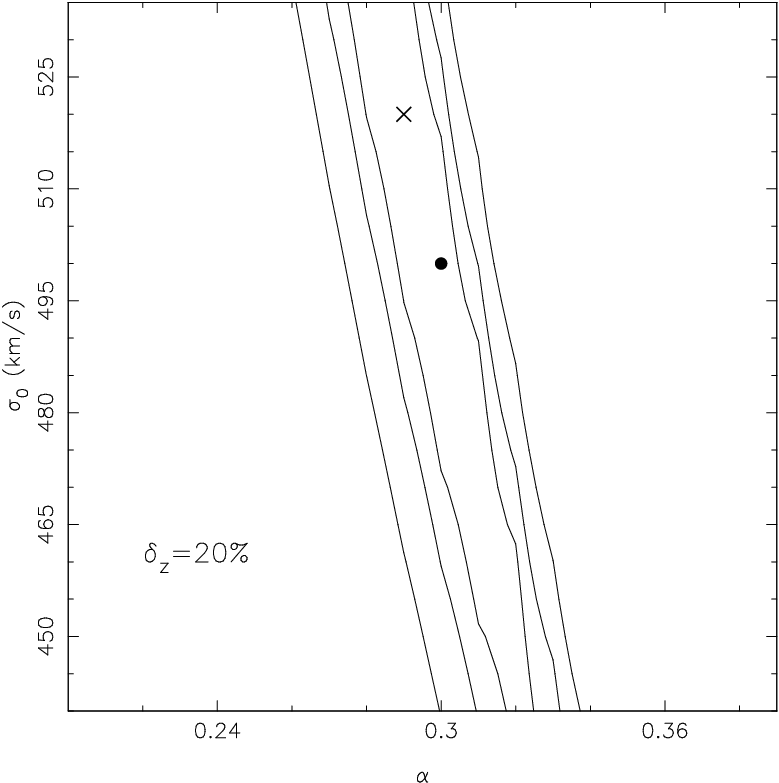}
\includegraphics[width=5.5cm,height=5.5cm]{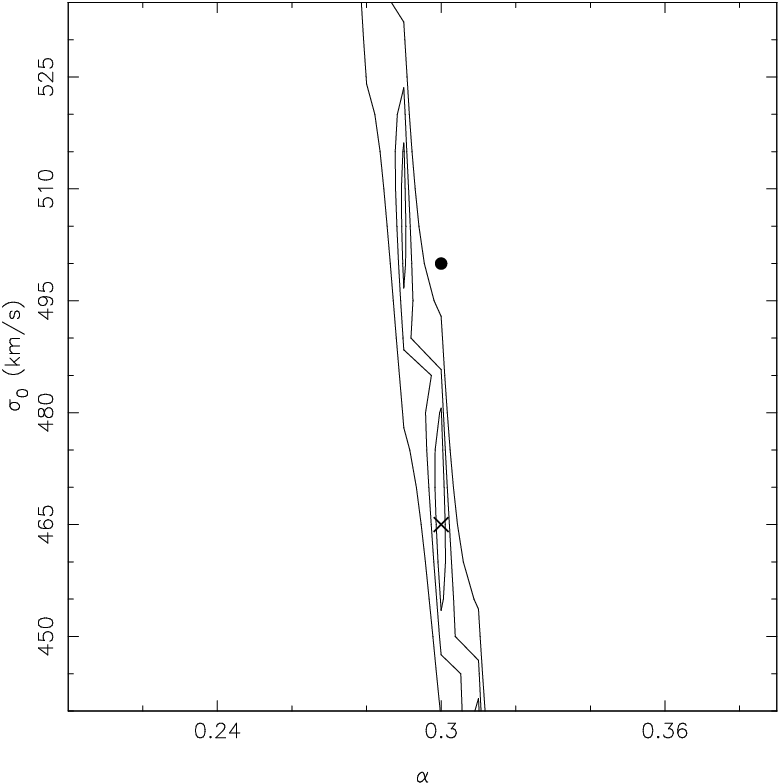}
\includegraphics[width=5.5cm,height=5.5cm]{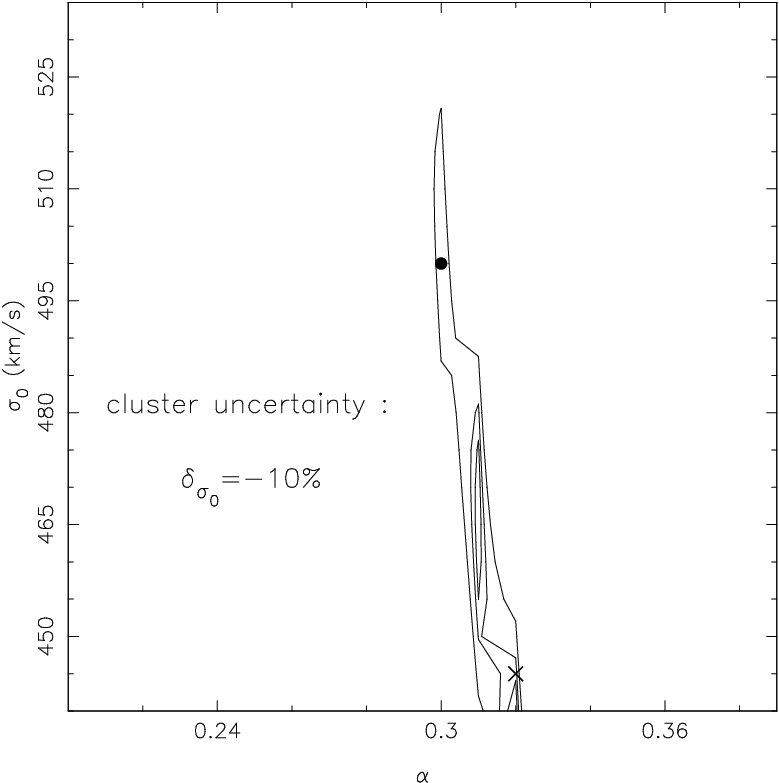}
\includegraphics[width=5.5cm,height=5.5cm]{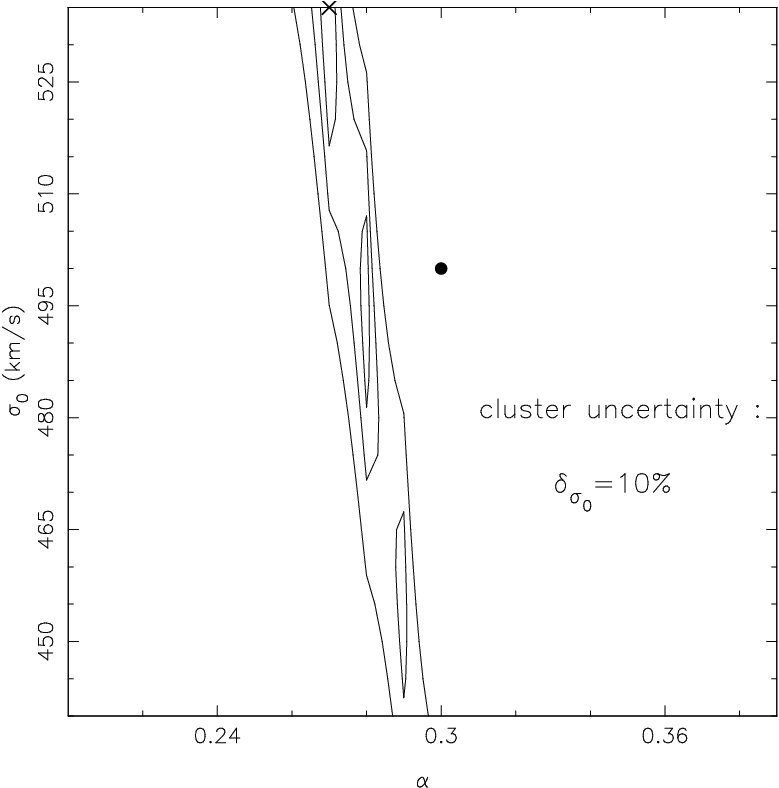}
\end{center}
\caption{\label{fig11} Results from the PL profile.
The top left plot (number 1) is the reference field situation with 25,000 elliptical background sources in a field of 26$\times$26 arcmin$^2$,
corresponding to a density of background sources equal to about 35 galaxies per arcmin$^2$.
Then the following plots (2, 3, 4) show the results obtained with 25,000 \emph{circular} sources,
then with 40,000 elliptical sources corresponding to a typical space
based observation yielding 60 galaxies per arcmin$^2$, then with
12,500 elliptical sources corresponding to ground based data with a
galaxy number density of about 17 galaxies per arcmin$^2$.
The plot number 5 and 6 show the effect of introducing an uncertainty in the
mean redshift of the source population: this uncertainty is equal
to -20\% (left) and +20\% (right).
The last row show the results from the cluster configuration:
the reference cluster configuration (plot 7), and then configurations
where uncertainty has been introduced in the cluster modeling. The
contours in all these panels represent the 3$\sigma$, 4$\sigma$ and 5$\sigma$
confidence levels. Note
  that the dot indicates input values and the cross the retrieved output.  } 
\end{figure*}

The sum is performed for a set of $r_{cut}$ values around the input
value used to simulate the catalogue and the range defined by the projection
of the 3$\sigma$ contour level along the $r_{cut}$ axis.
The results do not depend strongly on the range used to do the sum.
Fig.~\ref{fig7} and \ref{fig8} shows the $\mathcal{L'}$ contours we get for the NFW
and the PIEMD profiles. Since the PL profile does not have a cut-off radius
we cannot compare it easily with the re-parameterized PIEMD and NFW profiles.
The plots above show that we can put strong constraints on the aperture 
mass, the crossed line represents the line $M_{aper}(R_{aper})$ as 
computed with the input model used to generate the simulated catalogue.  

The motivation of such a re-parameterisation is that we deal with
more direct physical quantities than halo parameters, that is an aperture
mass calculated within an aperture radius. The primary motivation for
galaxy-galaxy lensing studies was to measure halo masses, so this
offers a more convenient parameterisation for that purpose.
This is also a different way of measuring masses compared to the aperture densitometry
method. Moreover, it is more suited to the case of clustered galaxies since
we are not able to integrate the shear profile for any individual galaxy.

To conclude in this paper, we have discussed galaxy-galaxy lensing in the context
of measuring masses of field and cluster galaxies. We compare the robustness of 
recovering input parameters for the mass distribution of lenses from a
ground based survey and space based observations.
We explore a wide range of input mass models for galaxy halos. We
simulate the galaxy-galaxy lensing effect and generate synthetic catalogues.
A maximum likelihood method is applied to the catalogues to successfully 
recover the lens parameters in various configurations. Going beyond
the standard parameterisation of a dark matter halo, we propose
a re-parameterisation of the problem in terms of more direct physical 
quantities: the aperture mass calculated within an aperture radius. 
The main result of this re-parameterisation is that we are able to put 
even stronger constraints on the aperture mass for an L$^*$ galaxy.

\section*{Acknowledgements}
ML thanks the Astrophysics Department of the Pontificia 
Universidad Cat\'olica, Santiago,
Chile for their hospitality, where a significant portion of
this work was completed.

\clearpage
\newpage

\newpage
\newpage

\label{lastpage}
\end{document}